

\font\titlefont = cmr10 scaled\magstep 4
 2
\font\sectionfont = cmr10
\font\littlefont = cmr5
\font\eightrm = cmr8 

\def\ss{\scriptstyle} 
\def\sss{\scriptscriptstyle} 

\newcount\tcflag
\tcflag = 0  

\ifnum\tcflag = 0 \magnification = 1200 \fi  

\global\baselineskip = 1.2\baselineskip
\global\parskip = 4pt plus 0.3pt
\global\abovedisplayskip = 18pt plus3pt minus9pt
\global\belowdisplayskip = 18pt plus3pt minus9pt
\global\abovedisplayshortskip = 6pt plus3pt
\global\belowdisplayshortskip = 6pt plus3pt


\def\endignore{}
\def\ignore #1\endignore{}

\newcount\dflag
\dflag = 0


\def\monthname{\ifcase\month
\or January \or February \or March \or April \or May \or June%
\or July \or August \or September \or October \or November %
\or December
\fi}

\newcount\dummy
\newcount\minute  
\newcount\hour
\newcount\localtime
\newcount\localday
\localtime = \time
\localday = \day

\def\advanceclock#1#2{ 
\dummy = #1
\multiply\dummy by 60
\advance\dummy by #2
\advance\localtime by \dummy
\ifnum\localtime > 1440 
\advance\localtime by -1440
\advance\localday by 1
\fi}

\def\settime{{\dummy = \localtime%
\divide\dummy by 60%
\hour = \dummy
\minute = \localtime%
\multiply\dummy by 60%
\advance\minute by -\dummy
\ifnum\minute < 10 
\xdef\spacer{0} 
\else \xdef\spacer{} 
\fi %
\ifnum\hour < 12 
\xdef\ampm{a.m.} 
\else 
\xdef\ampm{p.m.} 
\advance\hour by -12 %
\fi %
\ifnum\hour = 0 \hour = 12 \fi
\xdef\timestring{\number\hour : \spacer \number\minute%
\thinspace \ampm}}}



\def\endtitle{}
\def\title#1\endtitle{\vskip.5in\titlefont
\global\baselineskip = 2\baselineskip
#1\vskip.4in
\baselineskip = 0.5\baselineskip\rm}
 
\def\endauthors{}
\def\authors#1\endauthors{#1}

\def\endabstract{}
\def\abstract#1\endabstract{\vskip .3in%
\centerline{\sectionfont\bf Abstract}%
\vskip .1in
\noindent#1}

\def\nopageonenumber{\footline={\ifnum\pageno<2\hfil\else
\hss\tenrm\folio\hss\fi}}  

\newcount\nsection
\newcount\nsubsection

\def\section#1{\global\advance\nsection by 1
\nsubsection=0
\bigskip\noindent\centerline{\sectionfont \bf \number\nsection.\ #1}
\bigskip\rm\nobreak}

\def\subsection#1{\global\advance\nsubsection by 1
\bigskip\noindent\sectionfont \sl \number\nsection.\number\nsubsection)\
#1\bigskip\rm\nobreak}

\def\topic#1{{\medskip\noindent $\bullet$ \it #1:}} 
\def\endtopic{\medskip}

\def\appendix#1#2{\bigskip\noindent%
\centerline{\sectionfont \bf Appendix #1.\ #2}
\bigskip\rm\nobreak}


\newcount\nref
\global\nref = 1

\def\therefs{} 


\def\ref#1#2{\xdef #1{[\number\nref]}
\ifnum\nref = 1\global\xdef\therefs{\item{[\number\nref]} #2\ }
\else
\global\xdef\oldrefs{\therefs}
\global\xdef\therefs{\oldrefs\vskip.1in\item{[\number\nref]} #2\ }%
\fi%
\global\advance\nref by 1
}

\def\listrefs{\vfill\eject\section{References}\therefs}


\newcount\nfoot
\global\nfoot = 1

\def\foot#1#2{\xdef #1{(\number\nfoot)}
\footnote{${}^{\number\nfoot}$}{\eightrm #2}
\global\advance\nfoot by 1
}


\newcount\nfig
\global\nfig = 1
\def\thefigs{} 

\def\figure#1#2{\xdef #1{(\number\nfig)}
\ifnum\nfig = 1\global\xdef\thefigs{\item{(\number\nfig)} #2\ }
\else
\global\xdef\oldfigs{\thefigs}
\global\xdef\thefigs{\oldfigs\vskip.1in\item{(\number\nfig)} #2\ }%
\fi%
\global\advance\nfig by 1 } 

\def\figurecaptions{\vfill\eject\section{Figure Captions}\thefigs}

\def\fig#1{\xdef #1{(\number\nfig)}
\global\advance\nfig by 1 } 


\newcount\ntab
\global\ntab = 1

\def\table#1{\xdef #1{\number\ntab}
\global\advance\ntab by 1 } 


\newcount\cflag
\newcount\nequation
\global\nequation = 1
\def\eqlabel{(1)}

\def\nexteqno{\ifnum\cflag = 0
\global\advance\nequation by 1
\fi
\global\cflag = 0
\xdef\eqlabel{(\number\nequation)}}

\def\lasteqno{\global\advance\nequation by -1
\xdef\eqlabel{(\number\nequation)}}

\def\label#1{\xdef #1{(\number\nequation)}
\ifnum\dflag = 1
{\escapechar = -1
\xdef\draftname{\littlefont\string#1}}
\fi}

\def\clabel#1#2{\xdef\eqlabel{(\number\nequation #2)}
\global\cflag = 1
\xdef #1{\eqlabel}
\ifnum\dflag = 1
{\escapechar = -1
\xdef\draftname{\string#1}}
\fi}

\def\cclabel#1#2{\xdef\eqlabel{#2)}
\global\cflag = 1
\xdef #1{\eqlabel}
\ifnum\dflag = 1
{\escapechar = -1
\xdef\draftname{\string#1}}
\fi}


\def\eeq{}

\def\eqnn #1\eeq{$$ #1 $$}

\def\eq #1\eeq{
\ifnum\dflag = 0
{\xdef\draftname{\ }}
\fi 
$$ #1
\eqno{\eqlabel \rlap{\ \draftname}} $$
\nexteqno}







\def\eqa #1\eeq{
\ifnum\dflag = 0
{\xdef\draftname{\ }}
\fi 
$$ \eqalignno{ #1 } $$
\global\cflag = 0}


\def\ie{{\it i.e.\/}}
\def\eg{{\it e.g.\/}}
\def\etc{{\it etc.\/}}
\def\etal{{\it et.al.\/}}
\def\apriori{{\it a priori\/}}


\def\apj#1#2#3{{\it Ap.\ J.} {\bf #1}, (19#2) #3}

\def\mpla#1#2#3{{\it Mod.\ Phys.\ Lett.} {\bf A#1}, (19#2) #3}
\def\nci#1#2#3{{\it Nuovo Cimento} {\bf #1} (19#2) #3}
\def\npb#1#2#3{{\it Nucl.\ Phys.} {\bf B#1} (19#2) #3}
\def\plb#1#2#3{{\it Phys.\ Lett.} {\bf #1B} (19#2) #3}

\def\prc#1#2#3{{\it Phys.\ Rev.} {\bf C#1} (19#2) #3}
\def\prd#1#2#3{{\it Phys.\ Rev.} {\bf D#1} (19#2) #3}

\def\prep#1#2#3{{\it Phys.\ Rep.} {\bf C#1} (19#2) #3}
\def\prl#1#2#3{{\it Phys.\ Rev.\ Lett.} {\bf #1} (19#2) #3}

\def\sjnp#1#2#3#4#5#6{{\it Yad.\ Fiz.} {\bf #1} (19#2) #3
[{\it Sov.\ J.\ Nucl.\ Phys.} {\bf #4} (19#5) #6]}


\global\nulldelimiterspace = 0pt



\def\frac#1#2{{{#1} \over {#2}}\,}  
\def\hf{{1\over 2}}
\def\nth#1{{1\over #1}}


\def\Asl{\hbox{/\kern-.7500em\it A}} 
\def\Dsl{\hbox{/\kern-.6700em\it D}} 
\def\dsl{\hbox{/\kern-.5300em$\partial$}}
\def\pxpsl{\hbox{/\kern-.5600em$p$}}
\def\sslsh{\hbox{/\kern-.5300em$s$}}
\def\epssl{\hbox{/\kern-.5100em$\epsilon$}}
\def\delsl{\hbox{/\kern-.6300em$\nabla$}}
\def\lxpsl{\hbox{/\kern-.4300em$l$}}
\def\elxpsl{\hbox{/\kern-.4500em$\ell$}}
\def\kxpsl{\hbox{/\kern-.5100em$k$}}
\def\qxpsl{\hbox{/\kern-.5000em$q$}}
\def\sla#1{\raise.15ex\hbox{$/$}\kern-.57em #1}
\def\Pl{\gamma_{\sss L}}
\def\Pr{\gamma_{\sss R}}



\def\roughly#1{\mathrel{\raise.3ex\hbox{$#1$\kern-.75em\lower1ex\hbox{$\sim$}}}}
\def\lsim{\roughly<}
\def\gsim{\roughly>}

\def\ol#1{\overline{#1}}



\def\bfe{{\bf e}}

\def\bfk{{\bf k}}

\def\bfq{{\bf q}}
\def\bfr{{\bf r}}

\def\bfx{{\bf x}}



\def\Sca{{\cal A}}

\def\Scc{{\cal C}}
\def\Scd{{\cal D}}
\def\Sce{{\cal E}}
\def\Scf{{\cal F}}

\def\Sch{{\cal H}}

\def\Scl{{\cal L}}
\def\Scm{{\cal M}}
\def\Scn{{\cal N}}
\def\Sco{{\cal O}}

\def\Scr{{\cal R}}


\def\ssa{{\sss A}}
\def\ssb{{\sss B}}

\def\sse{{\sss E}}
\def\ssf{{\sss F}}

\def\ssj{{\sss J}}

\def\ssm{{\sss M}}
\def\ssn{{\sss N}}

\def\ssp{{\sss P}}

\def\sst{{\sss T}}

\def\ssv{{\sss V}}


\def\Tr{\mathop{\rm Tr}}

\def\Re{{\rm Re\;}}
\def\Im{{\rm Im\;}}
\def\diag#1{{\rm diag}\left( #1 \right)}


\def\bra#1{\langle #1 |}
\def\ket#1{| #1 \rangle}
\def\braket#1#2{\langle #1 | #2 \rangle}

\def\Avg#1{\left\langle #1 \right\rangle}




\def\eV{{\rm \ eV}}

\def\MeV{{\rm \ MeV}}

\input epsf.tex

 2
  
\overfullrule=0pt


\def\bk{\item{}}

\def\Asl{\hbox{/\kern-.7500em\it A}} 

\def\Avgb#1{\Avg{#1}_\ssb}
\def\Avgbp#1{\Avg{#1}_{\ssb'}}
\def\Avgce#1{\Avg{#1}_\Sce}
\def\Avge#1{\Avg{#1}_\sse}
\def\DU{\Delta U}

\def\ebar{\ol{e}}
\def\nbar{\ol{n}}
\def\pbar{\ol{p}}
\def\Bbar{\ol{B}}
\def\Ebar{\ol{E}}
\def\Ubar{\ol{U}}
\def\Vbar{\ol{V}}
\def\jbar{\ol{j}}
\def\nbar{\ol{n}}

\def\nubar{\ol{\nu}}

\def\sw{s_w}

\def\GF{G_{\sss F}}


\line{hep-ph/9606295 \hfill McGill-96/18}

\vskip .1in

\title
\centerline{Neutrino Propagation} 
\centerline{in a Fluctuating Sun}
\endtitle

\vskip .2in

\authors
\centerline{C.P. Burgess and D. Michaud}
\vskip .2in
\centerline{\it Physics Department, McGill University}
\centerline{\it 3600 University St., Montr\'eal, Qu\'ebec, Canada, H3A
2T8.}
\endauthors

\vskip .2in

\abstract
We adapt to neutrino physics a general formulation for particle propagation in
fluctuating media, initially developed for applications to electromagnetism and neutron
optics. In leading approximation this formalism leads to the usual MSW effective
hamiltonian governing neutrino propagation through a medium. Next-to-leading
contributions describe deviations from this description, which arise due to
neutrino interactions with fluctuations in the medium. We compute these corrections for
two types of fluctuations: ($i$) microscopic thermal fluctuations, and ($ii$) macroscopic
fluctuations in the medium's density. While the first of these reproduces standard
estimates, which are negligible for applications to solar neutrinos, we find the second
can be quite large, since it grows in size with the correlation length of the
fluctuation. We consider two models in some detail. For fluctuations
whose correlations are extend only over a local region in space of length $\ell$, 
appreciable effects for MSW oscillations arise if $(\delta n/n)^2 \ell
\gsim 100$ m or so. Alternatively, a crude model of helioseismic $p$-waves 
gives appreciable effects only when $(\delta n/n) \gsim 1\%$. In general
the dominant effect is to diminish the quality of the resonance, making the  
suppression of the ${}^7$Be neutrinos a good experimental probe of fluctuations
deep within the sun. Fluctuations can also provide a new  mechanism for 
reducing the solar neutrino flux, giving an energy-independent suppression factor
of $\hf$, away from the resonant region, even for small vacuum mixing angles.
\endabstract


\vfill\eject

\section{Introduction and Summary}

We do not understand our sun as well as we should. Although the sun shines brightly in
neutrinos,  it is not so bright as it ought to be according to our present understanding
of  its workings and of neutrino properties. 

\ref\SNExperiments{R. Davis, D.S. Harmer and K.C. Hoffman, \prl{20}{68}{1205};\bk
J.K. Rowley \etal, in {\it Solar Neutrinos and Neutrino Astronomy},
AIP Conference Proceedings number 126, edited by M.L. Cherry, W.A. Fowler
and K. Lande, (1985); \bk
K.S. Hirata \etal, \prl{65}{90}{1297}; \bk
P. Anselmann \etal, \plb{327}{94}{377}; \bk
J.N. Abdurashitov \etal, \plb{328}{94}{234}.}

\ref\SNTheory{The following references provide excellent reviews: \bk
J.N. Bahcall, {\it Neutrino Astrophysics}, Cambridge University Press, 1989; \bk
S. Turck-Chi\`eze \etal, \prc{230}{93}{57};\bk
W. Haxton, {\it Ann. Rev. of Astr. and Astrophys.} {\bf 33} (1995) 459.}

In recent years experimental and theoretical lines of research have converged to bring
this Solar Neutrino Problem (SNP) to a head.  On the one hand, confidence in the
experimentally-measured neutrino fluxes has grown with the observation of the
solar-neutrino shortfall in four independent experiments \SNExperiments, including
those which are capable of detecting neutrinos from the principal $p-p$ cycle of nuclear
reactions. Moreover, it has recently become possible to calibrate these detectors by
exposing them to very intense radioactive sources here on earth.  On the theoretical side,
confidence in the neutrino-flux predictions of solar models has also improved, for two
reasons \SNTheory. First, the redundancy of the experiments permits the discrepancy to
be mainly based on the predictions for the $p-p$ neutrinos. Since the $p-p$ reactions are
largely responsible for generating the sun's energy, the theoretical uncertainty in their
reaction rates is minimal. Second, the rise of the field of helioseismology has made
available an abundance of experimental data about the solar interior, thereby
significantly improving the constraints on the assumptions which must be made in
constructing solar models.

\ref\MSW{L. Wolfenstein, \prd{17}{78}{2369};\bk
V. Barger, K. Whisnant, S. Pakvasa and R.J.N. Phillips, \prd{22}{80}{2718};\bk
P. Langacker, J.P. Leville and J. Sheiman, \prd{27}{83}{1228};\bk
S.P. Mikheyev and A. Yu. Smirnov, {\it Sov. Phys. Usp.} {\bf 29} (1986) 1155;
\sjnp{42}{85}{1441}{42}{85}{913};
\nci{9}{86}{17};\bk
S.P. Rosen and J.M. Gelb, \prd{34}{86}{969};\bk
H. Bethe, \prl{56}{86}{1305};\bk
W. Haxton, \prl{57}{86}{1271};\bk
A.J. Baltz and J. Weneser, \prd{37}{88}{3364};\bk
P.D. Mannheim, \prd{37}{88}{1935}.}

If the problem is not the sun, then the measured neutrino shortfall must arise while
the neutrinos are {\it en route} to the earth. Besides gaining support from the
improvements in understanding of solar models, the credibility of such a neutrino
solution to the SNP is also boosted by the existence of a very plausible and elegant
mechanism for depleting the observed solar neutrino flux. The mechanism consists
of resonant (MSW) oscillations of the neutrinos as they pass through the sun \MSW. In
this picture the small influence of the solar medium on neutrino propagation plays an
important role by resonating with the equally small vacuum oscillations which
generically arise once neutrinos are endowed with masses. Considerable effort has been
invested in understanding the nature of these material-dependent oscillations.  

\ref\MeanOsc{A. Halprin, \prd{34}{86}{3462};\bk
A. Sch\"afer and S. Koonin, \plb{185}{87}{417};\bk
P. Krastev and A.Yu. Smirnov, \plb{226}{89}{341};\bk
P. Krastev and A.Yu. Smirnov, \mpla{~Vol. 6, No. 11}{91}{1001}; \bk
W. Haxton and W.-M. Zhang, \prd{43}{91}{2484}.}

\ref\Sawyer{R.F. Sawyer, \prd{42}{90}{3908}.}

\ref\Equilib{A. Dolgov, \sjnp{33}{81}{1309}{33}{81}{700};\bk
A. Manohar, \plb{186}{87}{370};\bk
L. Stodolsky, \prd{36}{87}{2273};\bk
R. Barbieri and A. Dolgov, \npb{349}{91}{743}.}

\ref\Nusinbath{S. Samuel, \prd{48}{93}{1462};\bk
V.A. Kostelecky, J. Pantaleone and S. Samuel, \plb{315}{93}{46};\bk
V.A. Kostelecky and S. Samuel, \prd{49}{94}{1740};\bk
J. Pantaleone, \plb{342}{95}{250};\bk
J.C.  d'Olivo and J.F. Nieves, preprint hep-ph/9501327.}

\ref\SiglRaffelt{G. Raffelt, G. Sigl and L. Stodolsky, \prl{70}{93}{2363};\bk
G. Sigl and G. Raffelt, \npb{406}{93}{423}.}

\ref\SolarAvging{For some influence of thermal averaging in the
sun, however, see: A. Loeb, \prd{39}{89}{1009};\bk
J. Rich, \prd{48}{93}{4318}.}

\ref\noisy{F.N. Loreti and A.B. Balantekin, preprint nucl-th/9406003;\bk
F.N. Loreti, Y.Z. Qian, G.M. Fuller and A.B. Balantekin, preprint astro-ph/9508106;\bk
E. Torrente Lujan, preprint hep-ph/9602398; \bk
A.B. Balantekin, J.M. Fetter and F.N.Loreti, preprint astro-ph/9604061.}

\ref\Denis{D. Michaud, McGill University M.Sc. thesis, 1994;\bk
in the 17th proceedings of the {\it Annual MRST Meeting}, 1995.}

\ref\noisytwo{H. Nunokawa, A. Rossi, V.B. Semikoz and J.W.F. Valle, 
preprint FTUV/95-47, IFIC/95-49, hep-ph/9602307.}

\ref\CMTexts{V.F. Sears, {\it Neutron Optics, An Introduction to the Theory
of Neutron Optical Phenomena and their Applications} (Oxford, 1989);\bk
C. Cohen-Tannoudji, J. Dupont-Roc and G. Grynberg, {\it Atom-Photons 
Interactions}, (Wiley, 1992).}

A common feature of the majority of these studies has been the approximation in
which the influence of the solar medium is described in terms of an effective
hamiltonian, depending on the mean values of the quantities to which the neutrinos
couple.  Less has been done to study scattering from the deviations away from
this mean. Some researchers have investigated the effects of 
neutrino scattering from position-dependent densities \MeanOsc, although usually
ignoring the potentially decohering effects \Sawyer\ --- more about which later 
--- of such scattering.  Incoherent scattering due to interactions with
the particles which make up the medium has also been studied within the 
context of supernovae, and the early universe, \Equilib\Nusinbath\SiglRaffelt,
in which case neutrinos themselves can be part of the underlying medium. This
type of scattering is entirely negligible within the sun \SolarAvging.
Until quite recently, \noisy\Denis\noisytwo, less attention has been devoted 
to the effects for solar neutrinos of more macroscopic fluctuations. 

The purpose of the present work is to develop a framework for describing the influence of
all such fluctuations on neutrino propagation, with the goal of identifying when each can be
important. To this end we adapt to neutrino physics a formalism which has been
extensively used to describe the interaction of electromagnetic waves and neutrons
with fluctuations in matter \CMTexts. As we describe in detail herein, our results agree with
earlier approaches when applied to the fluctuations they consider. 

In order of magnitude, fluctuation effects contribute to neutrino evolution
with strength $\GF^2 \Avg{\delta n \delta n} \ell_\|$. Here the average is
over different quantities in different situations, and $\delta n = n - \Avg{n}$ 
denotes the deviation of the particle density from its mean. $\ell_\|$ is the 
correlation length along the direction of neutrino motion. The relative
size of this term as compared to the usual MSW evolution term, $\GF \Avg{n}$,
is therefore of order $\GF \Avg{n} \ell_\| \epsilon^2$, where $\Avg{\delta n 
\delta n} = \epsilon^2 \Avg{n}^2$. Using the
central solar density, $\Avg{n} \sim 10^{26}/$cm${}^3$, we see that sizable effects
can be expected only for large-scale fluctuations: $\ell_\| \epsilon^2
\gsim (\GF \Avg{n} )^{-1} \sim 100$ km. Our more detailed analysis shows
that for resonant oscillations this estimate is too large, and sizable effects can
arise starting from $\epsilon^2 \ell_\| \gsim 100$ m. These scales
are of potential interest for solar neutrinos, since they are typical of scales
which can arise from physics within the sun. 

Some work on this kind of macroscopic-fluctuation-driven 
effects for neutrino propagation has appeared recently 
in the literature, starting with the pioneering work of ref.~\Sawyer. 
This reference considered a fixed density profile which varied in space, and 
computed the time evolution of the reduced density matrix which governs the
flavour degrees of freedom. Decohering fluctuation effects were found 
when the neutrino momenta were integrated out to obtain the flavour evolution.
We argue in Section 5 that this type of fluctuation is unlikely to be
important for solar-neutrino physics, although it could well play a role in
other applications. We reach this conclusion because we find that the only
fluctuations which can decohere neutrinos as they pass through, are those
whose size, $\ell_\perp$, transverse to the direction of neutrino propagation,
is smaller than the transverse size of the detector. But since fluctuations 
in the sun can in any case only affect neutrino evolution for $\ell_\| \gsim 100$ m, 
correlation lengths as small as typical neutrino detector sizes can play 
no significant role, so long as  $\ell_\perp \sim \ell_\|$. The same need not be
true for other applications, such as to supernovae, however.

More recently, the main approach to fluctuations that has been pursued to date, \eg\ by
refs.~\noisy\ and \noisytwo, is to model the density as a Gaussian random 
variable subject to the assumption that all correlations exist only over distances that
are negligibly small compared to the neutrino oscillation lengths which are of interest. Our
results extend these analyses in several ways. First, since we work from first
principles, we can give explicit expressions which are applicable to any
kind of density ensemble. In particular, we do {\it not} assume the correlation
length of the fluctuations to be small compared to neutrino oscillation lengths, and so
can apply our results to density profiles which vary on scales that are comparable to
the size of the sun. We can also incorporate arbitrary variation in space and 
time of the fluctuation's mean and variance. This permits us to consider 
such real density variations as helioseismic waves, which are known to exist 
in the sun. Our equations reduce to those of refs.~\noisy\ and \noisytwo\ in 
the limit that our assumptions overlap, but some of our most interesting applications
are to situations for which previous analyses do not apply. 

Our results are presented in the following way:

\topic{1}
We present a general formalism in Section 2 for describing the interactions of any
particle with arbitrary matter fluctuations. One of the main features of such
fluctuations is that they generically destroy the coherence of neutrino propagation by
evolving pure states into mixed states. As a result it is typically {\it impossible} to
describe them in terms of a matter-dependent effective hamiltonian, since any such 
hamiltonian would necessarily take pure states to pure states. This section
culminates in a master formula for the rate of change of the density matrix describing
particle propagation in an arbitrary medium, which naturally divides into a term which
defines a mean  effective matter-dependent hamiltonian, plus a fluctuation-dependent
term. A nice feature of the formalism is its recursive nature, which permits
fluctuations to be successively integrated out on larger and larger distance scales.

\topic{2}
Section 3 applies the general results of Section 2 to neutrinos moving in the presence
of microscopic fluctuations, which are those for which the correlation length is
negligible compared to the interesting scales for neutrino propagation. We consider
in some detail the special case of thermal fluctuations, and rederive the usual result
that these are small for neutrinos in the sun.

\topic{3}
Section 4 then considers matter fluctuations on larger scales. These fluctuations
arise because, although any one neutrino sees a fixed density profile, successive
neutrinos see different ones. We describe this by considering neutrinos to pass
through an ensemble of density profiles. The ensemble is characterized by 
expanding the density in terms of a complete set of 
modes whose amplitudes are taken to be uncorrelated random variables. The nature of
the underlying physics governs the basis of modes which are uncorrelated in any given
application. We consider two types of bases for illustrative purposes: ($i$)
fluctuations which are localized in position, within cells of slowly-varying length, $\ell$; and
($ii$) fluctuations in the amplitude of the normal modes which describe 
acoustic density waves. In the limit of small, constant, $\ell$ the first of 
these bases reduces to the case studied in refs.~\noisy\ and \noisytwo. The second 
is new, and is meant as a crude model of a helioseismic $p$-wave of fluctuating 
amplitude. We evaluate, for both examples, the
mean hamiltonian and the contributions of fluctuations to neutrino evolution.

\topic{4}
Section 5 takes the previous results and integrates out the neutrino momentum degrees of
freedom, to obtain the reduced evolution equation which governs the reduced
density matrix describing neutrino flavour and spin. 
We argue that previous derivations \Sawyer, which trace over 
neutrino momenta without taking into account that neutrino positions are 
ultimately measured in real experiments, give mistakenly large estimates of
the size of the decoherence which this trace introduces for solar neutrinos.  
A general expression is found for the fluctuation contributions to the 
neutrino-flavour evolution, which
are found to be characterized by a single parameter, $\Sca_{ab}$. This parameter is
evaluated for the two density ensembles introduced in Section 4. 

\topic{5}
Section 6 specializes the general results to the two flavour case, and integrates the
time evolution to obtain the electron-neutrino survival probability. 
An approximate analytic form
for this integration is obtained, which is a generalization of Parke's formula for
standard MSW mixing. The decoherence due to fluctuations appears in the
evolution as a damping term, similar (but not identical) to what would happen if the
neutrinos were decaying. Numerical integration is also performed, and found to agree well
with the analytical results. 

We perform the MSW analysis for both models of density fluctuations that are 
described in Section 4. It is found that appreciable changes
to the usual MSW scenario arise for surprisingly small amplitude fluctuations. In
terms of $\epsilon^2 = \Avg{\delta n \delta n}/\Avg{n}^2$, and the correlation length
$\ell_\|$, we  find deviations for $\epsilon^2 \ell_\| \gsim 10$ m.
Startlingly large changes from MSW behavior arise for $\epsilon^2 \ell_\| \gsim 1$ km. A
generic new feature of fluctuations is the introduction of a universal
energy-independent reduction of the survival probability to $\hf$ for 
small $E/\delta m^2$. 

The model of a fluctuating helioseismic $p$-wave gives discernible effects which are
comparatively small. For a wave with a 30 minute period, discernible effects require
$\epsilon \gsim 1 \%$. We understand the size of this effect to be due to the small
wavelength the wave typically has near the resonance region, due to the increase of the
speed of sound with depth in the medium. 
More realistic simulations are currently under way to see if the same
is true for neutrinos propagating through both solar $p$ and $g$ waves.

\topic{6}
Finally, in Section 7 the general formalism is applied to derive the effective
evolution equation governing the reduced neutrino flavour/spin density matrix
in the presence of a magnetic
moment interaction. Our conclusions are briefly summarized in Section 8.
\endtopic

\section{The General Formalism}

In order to keep all approximations explicit it is instructive to first formulate
our problem  within its most general context. Suppose, therefore, that our system 
consists of two sectors, $A$ and $B$, of which we wish to follow the evolution of
degrees of freedom in sector $A$ while ignoring (or partially ignoring -- see 
below) those in sector $B$. For example, when examining the influence of matter  
fluctuations on neutrino propagation we will take $A$ to describe the neutrino 
states of the system while $B$ consists of the states which are available to
the electrons and/or nucleons which make up the medium through which the 
neutrinos move. 

It is sometimes necessary to consider the more general case where a partial measurement
is made on sector $B$, in addition to the measurements which are performed in
sector $A$. We do this in order to set up the treatment of resonant oscillations,
for which $A$ consists only of the flavour (and spin) sectors of the 
single-particle neutrino sector, while $B$ contains both the neutrino 
position/momentum information as well as all medium-related effects. 
(We argue in Section 5 that an improper treatment of this case has 
in the past led to a mistaken estimate of the size of incoherent effects purely
due to this removal of momentum degrees of freedom.) 
The slightly more general formulation is required to analyze this situation  
since neutrino position information is in practice never completely ignored
(\ie\ neutrinos are all detected on Earth).
 
At an initial time, $t'$, we suppose these two sectors to be completely uncorrelated. 
That is, suppose the initial density matrix for the entire system factorizes:
\eq
\label\infact
\rho(t') = \varrho_\ssa \otimes \varrho_\ssb.
\eeq
We imagine here that $\varrho_\ssa$ acts only in the $A$ sector of the Hilbert
space and $\varrho_\ssb$ acts only in the $B$ sector, and we take $\varrho_\ssa$ and
$\varrho_\ssb$ to be separately normalized within their own sectors: 
$\Tr_\ssa \varrho_\ssa = \Tr_\ssb \varrho_\ssb = 1$. Here 
$\Tr_\ssa$ (or $\Tr_\ssb$) denotes a trace taken only over the 
$A$ (or $B$) sector of the Hilbert space. 

Next suppose the hamiltonian for the system takes the form:
\eq
\label\hamform
H = H_0 + \hat{V} ,
\eeq
in which $H_0 = H_\ssa + H_\ssb$ describes the separate evolution of the 
$A$ and $B$ sectors, while $\hat{V}$  is the interaction which couples these sectors 
together. Using this hamiltonian we may evolve $\rho$ to later times, $t > t'$. We
assume for this purpose that $\varrho_\ssa$ and $\varrho_\ssb$ respectively
commute with $H_\ssa$ and $H_\ssb$. Within the interaction representation 
the time evolution of $\rho(t)$ is then described by: 
\eq
\label\liouville
{\partial \rho \over \partial t} = -i \Bigl[ V(t) , \rho \Bigr],
\eeq
where $V(t) \equiv e^{i H_0 t} \, \hat{V} \, e^{-i H_0 t}$. Alternatively:
\eq
\label\intnevolution
\eqalign{
\rho(t) &= U(t,t') \, \rho(t') \, U^*(t,t') \cr
\hbox{with} \quad {\partial U(t,t') \over \partial t} &= -i V(t) \, U(t,t'). \cr}
\eeq
In general this time evolution will introduce correlations between
sectors $A$ and $B$ and so won't preserve the factorized form of eq.~\infact.
The remainder of this section is devoted to explicitly displaying these, and
other, effects as sector $A$ evolves in the presence of sector $B$.

\subsection{Coherent and Diffuse Scattering}

Suppose, now, that only observables associated with sector $A$ are to be measured at
some time $t > t'$. The probability of the results of any such measurement are
completely described by the reduced density matrix, $\rho_\ssa(t)$, 
defined by tracing the full density matrix over only the $B$ sector of states:
\eq
\label\reducedrho
\rho_\ssa(t) \equiv \Tr_\ssb \Bigl[ \rho(t) \Bigr].
\eeq
Notice that eq.~\infact\ implies $\rho_\ssa(t)$ satisfies the initial condition 
$\rho_\ssa(t=t') = \varrho_\ssa$. 

\vfill\eject

We now wish to split the time evolution for $\rho_\ssa(t)$ into a piece which
describes the mean features of sector $B$ --- `coherent' 
scattering\foot\coherence{We borrow the descriptions `coherent' and `diffuse' from 
the analogous applications of this formalism to the propagation of X-rays and neutrons
through matter. This split, as made precise in eq.~(8), is our $\ss definition$ 
of coherent scattering for the present purposes. Notice that the `coherent' part, 
as defined here, is coherent only in a weaker -- though more useful, for present
purposes --- sense than is sometimes used in electromagnetic applications. 
For instance, coherence in the present context need not imply  phase coherence
between incident and scattered waves.} --- plus a piece which describes 
the fluctuations about this mean --- `diffuse' scattering.   Our guiding principle
in so doing is to ensure that final time-evolved probabilities may be written 
as the non-interfering sum of a coherent part plus a diffuse part. 

Define, then, the mean (or coherent) evolution operator, $\Ubar(t,t')$, as the 
average of $U(t,t')$ over the $B$ sector, as follows:
\eq
\label\ubardef
\Ubar(t,t') \equiv \Avgb{U(t,t')},
\eeq
where the $B$-average of any quantity is defined by: $\Avgb{\cdots} \equiv 
\Tr_\ssb \Bigl[\varrho_\ssb ( \cdots ) \Bigr]$. The difference between $U(t,t')$
and $\Ubar(t,t')$ we denote:
\eq
\label\dudef
\DU(t,t') \equiv U(t,t') - \Ubar(t,t'),
\eeq
and so satisfies the defining identity $\Avgb{\DU} = 0$. 

With these definitions all probabilities calculated at times $t > t'$
are the sum of a coherent piece and a diffuse piece. That is, for any
hermitian observable acting only in the $A$ sector, $\Sco_\ssa$, eqs.~\infact,
\intnevolution, \ubardef\ and \dudef\ imply:
\eq
\label\obssum
\eqalign{
\Avg{\Sco_\ssa}(t) &\equiv \Tr \Bigl[ \rho(t) \Sco_\ssa \Bigr] \cr
&= \Tr \Bigl[ U(t,t') \rho(t') U^*(t,t') \Sco_\ssa \Bigr] \cr
&= \Tr_\ssa \Bigl[ \Ubar(t,t') \varrho_\ssa \Ubar^*(t,t') \Sco_\ssa \Bigr] 
+ \Tr \Bigl[ \DU(t,t') \rho(t') \DU^*(t,t') \Sco_\ssa \Bigr] \cr
&\equiv \Avg{\Sco_\ssa}_c(t) + \Avg{\Sco_\ssa}_d(t). \cr}
\eeq
The cross terms involving both $\Ubar(t,t')$ and $\DU(t,t')$ vanish by
virtue of the identity $\Avgb{\DU} = 0$. This last equality defines 
the mean (or coherent) and fluctuation (or diffuse) parts of 
$\Avg{\Sco_\ssa}(t)$. 

The distinction between diffuse and coherent evolution can also be made
directly for the reduced density matrix itself. That is, $\rho_\ssa(t) =
\rho_\ssa^c(t) + \rho_\ssa^d(t)$, where 
\eq
\label\dcsplitforrho
\eqalign{
\rho_\ssa^c(t) &\equiv \Ubar(t,t') \varrho_\ssa \Ubar^*(t,t'), \cr
\rho^d_\ssa(t) &\equiv \Tr_\ssb\Bigl[ \DU(t,t') \rho(t') 
\DU^*(t,t') \Bigr] . \cr}
\eeq

For many applications --- including the description of neutrino 
oscillations --- it is preferable to formulate the diffuse-coherent
split for $\partial \rho/\partial t$ rather than for the integrated 
evolution operator, $U(t,t')$. This is because it is often possible to
use perturbation theory for $\partial \rho/\partial t$ but not for
the long-time evolution of $\rho(t)$. This leads us to formulate the
main result of this section. The differential evolution equation 
for $\rho^c_\ssa(t)$ may be written:
\eq
\label\mastereqn
{\partial \rho^c_\ssa \over \partial t} = -i \Bigl[ \Vbar(t) \rho^c_\ssa(t) 
- \rho^c_\ssa(t) \Vbar^*(t) \Bigr], 
\eeq
where $\Vbar(t)$ is the {\it effective interaction hamiltonian} which
is defined in such a way as to ensure that $\Vbar$ is related to
$\Ubar(t,t')$ in the same way that $V$ is related to $U$. That is:
\eq
\label\vbardef
\Vbar(t) \equiv i \; {\partial \Ubar \over \partial t} \; \Ubar^{-1},
\eeq
which need {\it not} be hermitian (since $\Ubar$ need not be unitary). 

Similarly, the differential evolution equation for $\rho^d_\ssa(t)$ is 
\eq
\label\drhodiff
{\partial \rho^d_\ssa \over \partial t} =  {\partial \over \partial t} 
 \Tr_\ssb \Bigl[ \DU(t,t') \rho(t') \DU^*(t,t') \Bigr].
\eeq

\subsection{Incorporating a Partial Measurement in Sector $B$}

For some applications it is true that measurements do not completely ignore what
is going on in sector $B$. For instance, in applications to solar-neutrino oscillations 
we will follow common practice and take $A$ to describe only the neutrino flavour and 
spin degrees of freedom. This involves banishing all neutrino position and momentum
information into sector $B$, even though any realistic measurements do 
include some information concerning neutrino position, such as that they are 
detected on earth. This section describes the slight generalization of the formalism
which is required to handle such cases.  

We therefore relax the assumption that all of the observables of interest, $\Sco$,
need act only in sector $A$. Instead we assume them to involve a specific
observation in sector $B$, which is uncorrelated with all measurements in sector
$A$. That is, consider the class of observables having the form: 
\eq
\label\uncorobs
\Sco \equiv \Sco_\ssa \otimes \Sco_\ssb,
\eeq
with $\Sco_\ssa$ (or $\Sco_\ssb$) acting only in sector $A$ (or $B$). 
In this case expressions similar to those found above may be derived, in which
all averages over sector $B$ are weighted by the observable $\Sco_\ssb$.

Specifically, define once more the evolution operators $\Ubar(t,t')$ and $\DU(t,t')$
as in eqs.~\ubardef\ and \dudef, but with the $B$-average now defined 
by:\foot\adjoint{Beware: the operator ordering in this definition has the
counterintuitive implication that $\ss \Avgb{X}^*$ need $\ss not$ equal 
$\ss \Avgb{X^*}$.} 
\eq
\label\newbav
\Avgb{\cdots} \equiv {\Tr_\ssb \Bigl[ (\cdots) \varrho_\ssb \Sco_\ssb \Bigr]
\over \Tr_\ssb \Bigl[ \varrho_\ssb \Sco_\ssb \Bigr]}.
\eeq
This choice preserves the property that $\Avg{\Sco}(t)$ may be written as the
non-interfering sum of a diffuse and coherent contribution, although eq.~\obssum\
is slightly modified to become:
\eq
\label\newobssum
\eqalign{
\Avg{\Sco}(t) &\equiv \Tr \Bigl[ \rho(t) \Sco \Bigr] \cr
&= \Tr_\ssa \Bigl[ \Ubar(t,t') \varrho_\ssa \Ubar^*(t,t') \Sco_\ssa \Bigr] 
\; \Tr_\ssb \Bigl[\varrho_\ssb \Sco_\ssb \Bigr]
+ \Tr \Bigl[ \DU(t,t') \rho(t') \DU^*(t,t') \Sco \Bigr] \cr
&\equiv \Avg{\Sco}_c(t) + \Avg{\Sco}_d(t). \cr}
\eeq

As before, the time evolution of any such observable may be completely
described in terms of a reduced density matrix, $\rho_\ssa(t)$, for which
the definition, eq.~\reducedrho, is now replaced by:
\eq
\label\newreducedrho
\rho_\ssa(t) \equiv \Tr_\ssb \Bigl[ \rho(t) \Sco_\ssb \Bigr] ,
\eeq
satisfying the initial condition: $\rho_\ssa(t=t') = \varrho_\ssa \Tr_\ssb[
\varrho_\ssb \Sco_\ssb]$. Notice that $\rho_\ssa(t)$ defined this way is
{\it not} normalized. The differential time evolution of its coherent
part is now given by the analog of eq.~\mastereqn:
\eq
\label\newmastereqn
{\partial \rho^c_\ssa \over \partial t} = -i \Bigl[ \Vbar(t) \rho^c_\ssa(t) 
- \rho^c_\ssa(t) \Vbar^*(t) \Bigr] \; \Tr_\ssb \Bigl[ \varrho_\ssb \Sco_\ssb 
\Bigr] .
\eeq
Notice that this equation lacks a term proportional to ${\partial \Sco_\ssb \over 
\partial t}$, even though $\Sco_\ssb(t)$ generally depends on time within the
interaction picture with which we are working. Its omission from eq.~\newmastereqn\ 
is justified since there $\Sco_\ssb$ should be evaluated at the time, $t_m$, 
when the measurement is performed, rather than at the time, $t$, of the evolution.
$\Vbar(t)$ is once again defined by eq.~\vbardef.

The diffuse evolution now becomes:
\eq
\label\masterdiff
{\partial \rho^d_\ssa \over \partial t} \equiv  {\partial \over \partial t}
\Tr_\ssb \Bigl[ \DU(t,t') \rho(t') \DU^*(t,t') \Sco_\ssb \Bigr].
\eeq 

\subsection{Perturbative Expressions}

It is instructive to evaluate eqs.~\vbardef\ and \masterdiff\  
perturbatively in the interaction $V$. To this end we use the
familiar series solution to eq.~\intnevolution:
\eq
\label\ptbnforu
U(t,t') = \sum_{n=0}^\infty (-i)^n \, \int_{t'}^t d\tau_1 \cdots 
\int_{t'}^{\tau_{n-1}} d\tau_n
\; V(\tau_1) \cdots V(\tau_n) .
\eeq
Using this expression in the previous results leads to the following
formula for effective hamiltonian:
\eq
\label\vbarpthy
\Vbar(t) = \Avgb{V(t)} -i \int_{t'}^t d\tau \; \Avgb{ \delta V(t) \; 
\delta V(\tau) } + O\left(V^3\right),
\eeq
with $\delta V(t) \equiv V(t) - \Avgb{V(t)}$. 

Notice that the antihermitian part of $\Vbar$ first arises at second order in $V$.
For instance, if $[\varrho_\ssb, \Sco_\ssb] = 0$:
\eq
\label\reandimparts
\eqalign{ \hf \Bigl(\Vbar + \Vbar^* \Bigr) &= \Avgb{V} - \, {i \over 2} 
\int_{t'}^t d\tau \; \Avgb{ \Bigl[ \delta V(t), \delta V(\tau) \Bigr]} +
O\left(V^3\right), \cr
- \, {i \over 2} \, \Bigl(\Vbar - \Vbar^* \Bigr) &=  - \, {1 \over 2} 
\int_{t'}^t d\tau \; \Avgb{ \Bigl\{ \delta V(t), \delta V(\tau) \Bigr\} } +
O\left(V^3\right) \cr}
\eeq

Similarly, the rate of change of $\rho^d_\ssa$ is:
\eq
\label\diffpthy
{\partial \rho^d_\ssa \over \partial t} = \int_{t'}^t d\tau \; \Tr_\ssb \Bigl[ \Bigl(
 \delta V(t) \, \rho(t') \, \delta V(\tau)  +  \delta V(\tau) \, \rho(t') \, 
\delta V(t) \Bigr) \Sco_\ssb \Bigr]  + O\left(V^3\right).  
\eeq
Eqs.~\vbarpthy\ and \diffpthy\ are our starting point for applications of
this formalism to neutrino propagation through matter. 

\vfill\eject

\subsection{Long-Time Evolution and Master Equations}

Regardless of how small the interaction hamiltonian should be, perturbation
theory eventually fails if one follows the system's evolution for sufficiently 
long times. Worse, it is often precisely the long-time behaviour which is of
interest in particular applications. In this section we outline how to use the
above perturbative expressions for time scales which are sufficiently large
compared to the correlation times which govern the fluctuations in the medium. 

Suppose, then, that the correlation, $\Avg{\delta V(t) \delta V(t')}$, is
negligible for $|t-t'|$ greater than some correlation time, $\tau$. Suppose
also that the system's time evolution is required over timescales, $T$, for 
which $T \gg \tau$. In this case perturbative expressions for ${\partial
\rho_\ssa \over \partial t}$ may be useful provided that $\tau$ is small enough to
justify perturbation theory, even if the same would not be true for timescales
as large as $T$. 

In this limit a coarse-grained time derivative of the density matrix may be
defined for times which are large compared to $\tau$ \CMTexts. It is given by
neglecting the difference between $\rho_\ssa$ and $\rho^c_\ssa$ in the expressions for
${\partial \rho^c_\ssa \over \partial t}$, as well as neglecting the difference
between $\rho(t)$ and $\rho(t')$ in ${\partial \rho^d_\ssa \over \partial t}$.
Since these differences are higher order in the perturbation, $V$, this neglect
is justified over time scales which are short enough to lay within the domain of
perturbation theory. With these approximations, the perturbative expression
for the sum of eqs.~\mastereqn\ and \drhodiff\ becomes:
\eq
\label\markoveqn
\eqalign{
{\partial \rho_\ssa \over \partial t} &= {\partial \rho^c_\ssa \over \partial t}
+ {\partial \rho^d_\ssa \over \partial t}  \cr
&=  -i \Bigl[ \Vbar_2(t) \rho_\ssa(t) - \rho_\ssa(t) \Vbar_2^*(t) \Bigr] \cr
& \qquad \qquad +  \int_{t'}^t d\tau \; \Tr_\ssb \Bigl[ \Bigl( \delta V(t) \, \rho(t)
\, \delta V(\tau)  +  \delta V(\tau) \, \rho(t) \,  \delta V(t) \Bigr) \Sco_\ssb
\Bigr]  + O\left(V^3\right) ,\cr}
\eeq
where $\Vbar_2(t)$ denotes the second-order expression, eq.~\vbarpthy. 

If the correlation scale, $\tau$, is now assumed to be small compared to the time
scale over which the coarse graining is taken, then we may neglect correlations on
the right-hand-side of eq.~\markoveqn, by writing $\rho(t) \approx \rho_\ssa(t)
\otimes \rho_\ssb$. With this choice eq.~\markoveqn\ describes a Markov-like process,
for which ${\partial \rho_\ssa \over \partial t}$ depends only on $\rho_\ssa(t)$, and
not on the behaviour of $\rho_\ssa$ for times previous to $t$. This represents a great
simplification once eq.~\markoveqn\ is integrated to obtain the evolution of
$\rho_\ssa$ for very long times. It is this form of the time-evolution equations
which is used in Section 6 for describing neutrino evolution within the sun. 

\vfill\eject

\subsection{Unitarity and Decoherence}

There are two general features of the above expressions which 
bear special emphasis. 
Notice first that the condition $\Tr \rho = 1$ implies the same is
true for the reduced density matrix (when $\Sco_\ssb = I$): 
$\Tr_\ssa \rho_\ssa = 1$.
This is easily seen to follow from eqs.~\mastereqn, \vbarpthy\ and
\diffpthy\ by virtue of the following identity, which expresses the
optical theorem in the present example: 
\eq
\label\unitarity
{\partial \over \partial t} \Tr_\ssa \rho_\ssa = 
- i \Tr_\ssa \Bigl\{ \rho^c_\ssa(t) \Bigl[ \Vbar(t) - \Vbar^*(t) \Bigr] \Bigr\}  
+ \Tr_\ssa \; {\partial \rho^d_\ssa \over \partial t} = 0.  
\eeq

Second,
$\partial \rho^d_\ssa/\partial t$ and $\Vbar - \Vbar^*$ both cause a loss of coherence
within sector $A$. That is, if the system is initially prepared in a pure
state, for which $\rho_\ssa^2 = \rho_\ssa$, then it need not remain so
after interacting with sector $B$. The endpoint of evolution is therefore
generally a mixed state. Quantitatively, starting from an initially pure
state eqs.~\mastereqn\ and \drhodiff\ imply the following rate of coherence loss:
\eq
\label\decoherence
\left. {\partial \over \partial t} \; \Bigl( \rho_\ssa^2 - \rho_\ssa \Bigr) 
\right|_{\rho_\ssa^2 = \rho_\ssa} = -i \; \rho_\ssa \Bigl( \Vbar - 
\Vbar^* \Bigr) \rho_\ssa + \left\{ \rho_\ssa , {\partial \rho^d_\ssa \over 
\partial t} \right\} - {\partial \rho^d_\ssa \over \partial t} , 
\eeq
where we have neglected the difference between $\rho_\ssa$ and $\rho_\ssa^c$
in the first term on the right-hand-side. 
When this is nonzero it clearly makes no sense to define the time evolution
in terms of the Schr\"odinger evolution, $ i {\partial \over \partial t} 
\ket\psi = H \ket\psi$, for a pure state: $\rho_\ssa = \ket\psi \, \bra\psi$. 

\subsection{Recursiveness}

Notice that the definitions of the effective hamiltonian, $\Vbar$, and 
of the diffuse scattering term, $\partial\rho^d_\ssa/\partial t$, 
are recursive, in the following sense.
Suppose that sector $A$ in the previous discussion were itself to be divided
into independent subsectors, $A'$ and $B'$, and only observables acting
in subsector $A'$ were measured. Suppose also that the initial state
did not involve any correlations between these two subsectors: $\varrho_\ssa
= \varrho_{\ssa'} \otimes \varrho_{\ssb'}$. Then we may further 
reduce the density matrix to act only within this subsector:
\eq
\label\newrhodef
\rho_{\ssa'} (t) \equiv \Tr_{\ssb'} \Bigl[ \rho_\ssa(t) \Bigr] = 
\Tr_{\ssb' \cup \ssb} \Bigl[ \rho(t) \Bigr].
\eeq

Then we define the coherent and diffuse part of the evolution in sector $A'$
in such a way as to ensure that the coherent part takes the same form 
regardless of whether the trace over sectors $B$ and $B'$ are performed 
separately, or all at once. That is, with the definitions 
\eq
\label\recubardefs
\Ubar_\ssb \equiv \Tr_\ssb\Bigl[ \varrho_\ssb U \Bigr] \qquad
\hbox{and} \qquad \Ubar_{\ssb'} \equiv \Tr_{\ssb'} \Bigl[ 
\varrho_{\ssb'} \Ubar_\ssb \Bigr] = \Tr_{\ssb \cup \ssb'} \Bigl[ 
(\varrho_\ssb \otimes \varrho_{\ssb'}) U \Bigr] ,
\eeq
and
\eq
\label\recdudefs
\DU_\ssb \equiv U - \Ubar_\ssb \qquad
\hbox{and} \qquad \DU_{\ssb'} \equiv \Ubar_\ssb - \Ubar_{\ssb'} ,
\eeq
we have $\Tr_\ssb \Bigl[ \varrho_\ssb \DU_\ssb \Bigr] = 
\Tr_{\ssb'} \Bigl[ \varrho_{\ssb'} \DU_{\ssb'} \Bigr] = 0$, and so the
expectation of any observable in only sector $A'$ may be written:
\eq
\label\recobssum
\eqalign{
\Avg{\Sco_{\ssa'}}(t) &\equiv \Tr \Bigl[ \rho(t) \Sco_{\ssa'} \Bigr] \cr
&= \Tr_\ssa \Bigl[ \Ubar_\ssb(t,t') \varrho_\ssa \Ubar_\ssb^*(t,t') \Sco_{\ssa'} \Bigr] 
+ \Tr \Bigl[ \DU_\ssb(t,t') \rho(t') \DU_\ssb^*(t,t') \Sco_{\ssa'} \Bigr] \cr
&= \Tr_{\ssa'} \Bigl[ \Ubar_{\ssb'}(t,t') \varrho_{\ssa'} \Ubar_{\ssb'}^*(t,t') 
\Sco_{\ssa'} \Bigr] 
+ \Tr_\ssa \Bigl[ \DU_{\ssb'}(t,t') \varrho_{\ssb'} \DU_{\ssb'}^*(t,t') 
\Sco_{\ssa'} \Bigr] \cr
& \qquad \qquad \qquad \qquad + \Tr \Bigl[ \DU_\ssb(t,t') \rho(t') 
\DU_\ssb^*(t,t') \Sco_{\ssa'} \Bigr] \cr}
\eeq

A similar expression holds for ${\partial \rho_{\ssa'} \over \partial t}$:
\eq
\label\recmastereqn
{\partial \rho_{\ssa'} \over \partial t} = -i \Bigl[ \Vbar_{\ssb'}(t) 
\rho_{\ssa'}(t) - \rho_{\ssa'}(t) \Vbar_{\ssb'}^*(t) \Bigr] + {\partial 
\rho^d_{\ssb'} \over \partial t} + {\partial \rho^d_\ssb \over \partial t} ,
\eeq
where:
\eq
\label\recmasterdiff
\eqalign{
\rho^d_{\ssb'} & \equiv \Tr_{\ssb'} \Bigl[ \DU_{\ssb'}(t,t') \rho(t')
 \DU_{\ssb'}^*(t,t') \Bigr] \cr
\rho^d_{\ssb} &\equiv \Tr_{\ssb' \cup \ssb} \Bigl[\DU_\ssb(t,t') 
\rho(t') \DU_\ssb^*(t,t') \Bigr].
\cr}
\eeq

The effective hamiltonians, $\Vbar_\ssb(t)$ and $\Vbar_{\ssb'}(t)$, are 
respectively defined, as usual, in terms of $\Ubar_\ssb(t,t')$ and
$\Ubar_{\ssb'}(t,t')$, using eq.~\vbardef. In particular, using the
notation $\Avgbp{\cdots} = \Tr_{\ssb'}[(\cdots) \varrho_{\ssb'} ]$,
we have the following very useful perturbative expression:
\eq
\label\recptbnvtwo
\eqalign{
\Vbar_{\ssb'}(t) &= \Avgbp{\Avgb{V(t)}} -i \int_{t'}^t d\tau 
\Bigl\{ \Bigl[ \Avgbp{\Avgb{ V(t) V(\tau)} - \Avgb{ V(t)} 
\Avgb{ V(\tau)}} \Bigr] \cr
& \qquad \qquad + \Bigl[ \Avgbp{ \Avgb{ V(t)}\Avgb{ V(\tau)}} 
- \Avgbp{\Avgb{ V(t)}}\Avgbp{ \Avgb{ V(\tau)}} \Bigr] \Bigr\} , \cr
&= \Avgbp{\Avgb{V(t)}} -i \int_{t'}^t d\tau 
\Bigl\{ \Avgbp{\Avgb{ V(t) V(\tau)}} - \Avgbp{\Avgb{ V(t)}}\Avgbp{ 
\Avgb{ V(\tau)}}  \Bigr\} . \cr}
\eeq

The recursive nature of these definitions is a very attractive feature. 
This is because
it lends itself to a renormalization-group-like analysis of
the effects of a medium on particle propagation, in which the effects of 
fluctuations on successively larger distance scales are separately integrated out.

\subsection{Neutrino Interactions}

Our later applications use the formalism just presented for the special
case of neutrinos interacting with matter through the weak interactions. 
(Magnetic moment interactions are briefly considered in Section 7.) 
We therefore pause here to gather the relevant expressions for the interaction
hamiltonian. 
None of the details of the nature of the medium are needed at this point, although
we suppose for simplicity that it does not include a significant component of 
neutrinos themselves.\foot\nusinbath{Although more than adequate for applications 
to the sun, this assumption can break down for supernovae or the early universe.}
This permits a clean separation between the neutrino states and the states
which are available to the medium. 

The neutrinos may be described by $N_\nu$ majorana neutrino fields, $\nu_i, i =
1,\dots,N_\nu$, without loss of generality.  In the absence of light sterile neutrinos 
this consists of the usual ($N_\nu =3$) neutrino eigenstates. The coupling
between neutrinos and the medium is mediated by the weak interactions.
To keep as broad as possible the applications of this section we choose:
\eq
\label\nuintlagr
\Scl = i \, \nubar_i \gamma_\mu \left( \Pl g^a_{ij}  + \Pr h^a_{ij} \right)
\nu_j \; J^\mu_a,
\eeq
where $\Pl$ and $\Pr$ project onto left- and right-handed spinors; $J_a^\mu$
are a set of hermitian operators involving the degrees of 
freedom of the medium, and $g^a_{ij}$ and $h^a_{ij}$ are corresponding 
$N_\nu \times N_\nu$ matrices of couplings. The reality of $\Scl$ implies $g^a_{ij}$ 
and $h^a_{ij}$ must all be hermitian.\foot\conventions{Our conventions are
$\ss \Pl = \hf(1 + \gamma_5)$,  $\ss \Pr = \hf(1 - \gamma_5)$ and 
$\ss \nubar = i \nu^\dagger\gamma^0 = -i \nu^\dagger \gamma_0$.} 

Since the most important applications are to the Standard Model (SM),
possibly supplemented by various sterile neutrinos and/or neutrino masses,
we record explicit expressions for the quantities, $J_a^\mu$, $g^a_{ij}$
and $h^a_{ij}$ in this case. The couplings to charged leptons, $\ell_m$,
are given by:
\eq
\label\SMleptcouplings
\eqalign{
(J^\mu_\pm)_{mn} &= i \ol{\ell}_m \gamma^\mu (1 \pm \gamma_5) \, \ell_n , \cr
\hbox{with} \quad (g^{mn}_+)_{ij} &= \sqrt2 \GF \left[ V_{mj} V^*_{ni}
+ \Scn_{ij} \left(- \, \hf + \sw^2 \right) \right], \cr
(g^{mn}_-)_{ij} &= \sqrt2 \GF \Scn_{ij}\; \sw^2 , \cr
\hbox{and} \quad (h^{mn}_\pm)_{ij} &= 0, \cr}
\eeq
where $V_{mi}$ are the leptonic CKM matrix elements which arise in the
charged-current couplings once neutrinos acquire masses, and $\Scn_{ij}$ 
are the analogous matrices which can arise in the neutrino neutral-current
couplings. $\sw$ denotes, as usual, the weak mixing angle. The corresponding
couplings to hadrons arise through their quark content:
\eq
\label\SMquarkcouplings
\eqalign{
(J^\mu_\pm)_a &= i \ol{q}_a \gamma^\mu (1 \pm \gamma_5) \, q_a , \cr
\hbox{with} \quad (g^{a}_+)_{ij} &= \sqrt2 \GF \Scn_{ij} \left(T_{3a} - Q_a
\sw^2 \right) , \cr 
(g^{a}_-)_{ij} &= \sqrt2 \GF \Scn_{ij} \left(-  Q_a \sw^2 \right) ,
\cr \hbox{and} \quad (h^{a}_\pm)_{ij} &= 0 ,\cr}
\eeq
where $T_{3a}$ and $Q_a$ are the third component of weak isospin and
electric charge of the corresponding quark. 

For many practical applications the energies involved are sufficiently low
that the relevant hadronic degrees of freedom are just protons and neutrons. 
In this case we may approximate eqs.~\SMquarkcouplings\
by the following effective macroscopic currents:
\eq
\label\SMnuclcouplings
\eqalign{
(g^a_+)_{ij} \; (J^\mu_+)_a + (g^a_-)_{ij} \; (J^\mu_-)_a  &\approx 
{\GF \over \sqrt2 } \, \Scn_{ij} \left[ \left( 1 - 4 \sw^2 \right) \; 
i \ol{p} \gamma^\mu p  -  i \ol{n} \gamma^\mu n \right] \cr
& \qquad\qquad + (\hbox{axial-current and higher-derivative terms}). \cr}
\eeq

We now use these expressions to compute the quantities $\Vbar$ and
${\partial \rho^d \over \partial t}$ in two different regimes. We first 
consider the case for which the fluctuations of interest occur on scales
which are microscopic in comparison to those relevant to neutrino 
propagation. This is followed, in Section 4, by a consideration of 
macroscopic fluctuations, for which correlation lengths are much larger.

\section{Microscopic Fluctuations}

Consider first the case of fluctuations having microscopic characteristic
correlation lengths. For neutrinos in the sun this includes the thermal 
fluctuations among the particles making up the solar interior, and so the
expressions obtained in this section may in this case be tested against standard
results.

Our goal is to compute the quantities $\Vbar$ and ${\partial \rho^d_\ssa \over \partial t}$
of Section 2. In order to apply the formalism we must first split the system into
sectors, $A$ and $B$. We choose $A=\nu$ to consist of all of the states in
the neutrino sector, while sector $B=E$ (`environment') represents the sector 
describing the other particles --- electrons, nucleons \etc\ --- through which 
the neutrinos propagate.\foot\labelchange{Our use of subscripts `$\ss \nu$' and
`$\ss E$', in place of `$\ss A$' and `$\ss B$', is meant to 
avoid confusion with the different choice for $\ss A$ and $\ss B$ which is
made in the subsequent sections.}

At first order in the interaction, eq.~\nuintlagr, we have
$\DU = 0$ and:
\eq
\label\thvbarone
\Vbar_1 = \Avge{V} = - \int d^3x \; i \, \nubar_i \gamma_\mu \left( \Pl g^a_{ij}  
+ \Pr h^a_{ij} \right) \nu_j \; \Avge{J^\mu_a}. 
\eeq
To this order neutrino evolution is simply described by replacing
the interaction current, $J_a^\mu$, with its mean, 
$\Avge{J_a^\mu} = \Tr_\sse \Bigl[\varrho_\sse J^\mu_a \Bigr]$, evaluated
using $\varrho_\sse (= \varrho_\ssb)$, which describes the initial
state of the medium through which the neutrinos pass. 

More can be said about the mean currents, $j^\mu_a(x) \equiv \Avge{J^\mu_a(x)}$,
given more information about the state $\varrho_\sse$. With the sun in mind we
take this to describe a mixture of nonrelativistic electrons, protons and neutrons
which are mutually interacting dominantly through the electromagnetic and strong
interactions. This permits the use of parity invariance to limit the form taken
by the mean currents. Moreover, we also work in an electroweak basis, 
for which $\Scn_{ij} = \delta_{ij}$ 
and $V_{mi} = \delta_{mi}$ for the three usual neutrinos, and $\Scn_{ij} = 
V_{mi} = 0$ for any light sterile neutrinos. (The influence of neutrino masses
is described in more detail once the trace over neutrino momentum states is
performed in Section 5.)
With these choices we obtain the usual estimate 
for the mean matter currents. We find $\Avge{ h^a_{ij} J^\mu_a} = 0$, and: 
\eq
\label\nbarests
\Avge{ g^a_{ij} \; J^\mu_a} \approx {\GF \over \sqrt2} \; 
\Bigl\{ 2  \delta_{ie} \, \delta_{je} \; j^\mu_e(x) - \delta_{ij} j^\mu_n(x) 
+ \delta_{ij} \, (1 - 4 \sw^2) \; [j^\mu_p(x) - j^\mu_e(x) ] \Bigr\} ,
\eeq
where $j^\mu_e = i \ebar \gamma^\mu e$, $j^\mu_p = i \pbar \gamma^\mu p$ and 
$j^\mu_n = i\nbar \gamma^\mu n$ are respectively the local 
electron, proton and neutron currents. This expression is obtained by 
averaging eqs.~\SMleptcouplings\ and \SMquarkcouplings, using eq.~\SMnuclcouplings. 
The axial-vector parts of the weak currents drop out of eq.~\nbarests\ 
by virtue of parity invariance of the solar medium. 
Notice that, although the neutron current distribution is independent of
the others, for slowly moving particles local electric neutrality implies $j^\mu_e(x) =
j^\mu_p(x)$,  and so the terms in the last equation which are proportional to 
$(1 - 4\sw^2)$ cancel. For nonrelativistic particles there is also a further
simplification, since we may neglect of all of the spatial components of the mean
currents: $j^\mu_a \approx n_a \delta^\mu_0$. 

At second order in the weak interactions two new things happen. First,
$\Vbar$ acquires a second-order correction, typically introducing to it an 
antihermitian part. Second, ${\partial \rho^d_\nu \over \partial t}$
becomes nonzero, introducing decoherence into any propagating neutrino state. 
We find:
\eq
\label\thvbartwo
\eqalign{
\Vbar_2 &= i  \int_{t'}^t d\tau \, d^3x \, d^3x' \; \nubar_i \gamma_\mu 
\left( \Pl g^a_{ij}  + \Pr h^a_{ij} \right) \nu_j \cr
& \qquad\qquad\qquad\qquad  \nubar_k' \gamma_\lambda \left( \Pl g^b_{kl}  + 
\Pr h^b_{kl} \right) \nu_l' \; \Avge{\delta J^\mu_a \delta J^\lambda_b{}'} , \cr}
\eeq
and
\eq
\label\thdiff
\eqalign{
\left({\partial \rho^d_\nu \over \partial t}\right)_2 &= - \int_{t'}^t d\tau 
\, d^3x \, d^3x' \; \Bigl\{ \nubar_i \gamma_\mu  \left( \Pl g^a_{ij}  
+  \Pr h^a_{ij} \right) \nu_j \cr
& \qquad\qquad  \rho_\nu \; \nubar_k' \gamma_\lambda   
\left( \Pl g^b_{kl} + \Pr h^b_{kl} \right) \nu_l' \;  \Avge{
\delta J^\lambda_b{}' \delta J^\mu_a } + 
(t \leftrightarrow \tau) \Bigr\}.    \cr}
\eeq
In these expressions $\rho_\nu ( = \rho_\ssa)$ denotes the neutrino density matrix at
time $t$. A prime on any field denotes that it is evaluated at the spacetime 
point $(\bfx',\tau)$ --- \eg\ $\nu' = \nu(\bfx', \tau)$ --- while all unprimed 
fields are evaluated at $(\bfx,t)$. 

At this point we use the information that the scale of the fluctuations, $\ell$, are 
microscopic in comparison with the distances of interest for neutrino propagation. 
This means the correlations, $\Avge{\delta J^\mu_a  \delta J^\lambda_b{}'}$, 
can be written in the approximate form 
\eq
\label\jjdeltaapprox
 \Avge{ \delta J^\mu_a \; \delta J^\lambda_b{}'} \approx 
C_{ab}^{\mu\lambda} \; \delta^3(\bfx - \bfx')  ,  
\eeq
where the coefficient functions, $C_{ab}^{\mu\lambda}$, are
explicitly calculable given the state $\varrho_\sse$ --- a point to which
we return below. 

Eq.~\jjdeltaapprox\ permits eqs.~\thvbartwo\ and \thdiff, for $\Vbar_2$ and
$\left({\partial \rho^d_\nu \over \partial t}\right)_2$, to be written as 
follows:
\eq
\label\lcvbartwo
\Vbar_2  = i  \int_{t'}^t d\tau \, d^3x \; C_{ab}^{\mu\lambda} \; \nubar_i \gamma_\mu 
\left( \Pl g^a_{ij}  + \Pr h^a_{ij} \right) \nu_j  \;\; \nubar_k' \gamma_\lambda 
\left( \Pl g^b_{kl} + \Pr h^b_{kl} \right) \nu_l' ,
\eeq
and
\eq
\label\lcdiff
\left({\partial \rho^d_\nu \over \partial t}\right)_2  = -2 \int_{t'}^t d\tau \, d^3x \; 
C_{ba}^{\lambda\mu} \; \nubar_i \gamma_\mu  
\left( \Pl g^a_{ij}   +  \Pr h^a_{ij} \right) \nu_j \; \rho_\nu \; \nubar_k'
\gamma_\lambda \left( \Pl g^b_{kl} + \Pr h^b_{kl} \right) \nu_l'.
\eeq

\ref\Forster{See, for example, a concise summary in:
D. Forster, {\it Hydrodynamic fluctuations, Broken Symmetry,
and Correlation Functions}, (Benjamin-Cummings, Reading Mass., 1980).}

Notice that these interactions describe processes, such as
$\nu \nu \to \nu\nu$ or $\nu \to \nu\nu\nu$, in which neutrinos scatter from 
medium-dependent fluctuations. Similar interactions are familiar for 
electromagnetic propagation  through matter, where the analogs of
eqs.~\lcvbartwo\ and \lcdiff\ are quadratic in the electromagnetic field
and so describe the scattering of electromagnetic waves by microscopic
fluctuations \Forster. (Similar quadratic terms also arise for neutrinos
at second order in their charged-current interactions with the 
particles in the bath.) Since the
matter-dependent effects depend differently on neutrino energies 
than do the same processes {\it in vacuo}  they can, in principle,  be separated 
from one another. 

The potential size of the matter-dependent correlation
coefficients, $C_{ab}^{\mu\lambda}$, may be estimated by computing 
them for thermal fluctuations in a system of nonrelativistic particles in 
local thermal equilibrium\foot\notcritical{This application presumes
not being near a critical point for which thermal fluctuations need not 
be microscopic in size.} and for which, for simplicity, we imagine there
is a single conserved particle number, $\Scn = \int d^3x \; J^0$, to whose
current the neutrinos couple.  

\ref\Kreuzer{A clear discussion of these issues may be found in:
H.J. Kreuzer, {\it Nonequilibrium Thermodynamics and its Statistical 
Foundations}, (Oxford, 1981).}

For thermal fluctuations we compute all averages over $E$ using the 
density matrix, $\varrho_\sse$, of the grand canonical ensemble:
\eq
\label\thermalstate
\varrho_\sse = Z^{-1} \; e^{ -  (H_\sse - \mu \Scn) / T}  .
\eeq
$Z$ here is the standard normalization constant: $Z = \Tr_\sse 
e^{- {(H_\sse - \mu \Scn)/ T}}$. 
More generally, we consider media which are
in local thermal equilibrium, and so for which $\varrho_\sse$ has 
a similar form, but with mean thermodynamic properties which vary
(over macroscopic distances) from place to place.  
The grand canonical ensemble is the one which is locally appropriate for 
this case \Kreuzer, since the number of particles in any local region of 
the medium is not fixed.

\ref\Huang{K. Huang, {\it Statistical Mechanics, 2nd Edition} 
(Wiley, New York, NY, 1987).}

With these choices we may compute the local fluctuations of $\Scn$. 
Using the assumption that the constituents of the medium are nonrelativistic, 
we neglect all but the time component of the current: $\Avge{J^\mu(x)} = n(x) \,
\delta^\mu_0$. We find \Huang: 
\eq
\label\thermalresult
C^{\mu\lambda} = \delta^\mu_0 \; \delta^\lambda_0 \Bigl[ n^2 
\kappa_\sst \, T + f_\ell(x) \Bigr],
\eeq
where $T(x)$ is the local temperature, and $\kappa_\sst(x)$ is the system's specific
isothermal compressibility: $\kappa_\sst = {1 \over n} \, \left( 
{\partial n \over \partial p} \right)_\sst$, where $p$ is the pressure. 

In eq.~\thermalresult, $f_\ell(x)$ is a function whose scale of variation is the
microscopic fluctuation length, $\ell$, and which satisfies the defining
condition: $\int d^3x \; f_\ell(x) = 0$. Because of these conditions, $f_\ell(x)$ 
can be neglected for macroscopic applications, such as when eq.~\thermalresult\
is used in eqs.~\lcvbartwo\ and \lcdiff\ to describe  
neutrino evolution over scales which are much larger than $\ell$. 

Substituting eq.~\thermalresult\ into expressions \lcvbartwo\ and \lcdiff, and
using the results in eq.~\mastereqn, reproduces the usual expressions 
\Equilib\Nusinbath\SiglRaffelt\  for 
neutrino scattering from a thermal ensemble of particles. This can be seen by
using the ideal-gas equation of state, $p = nT$, in which case
$\kappa_\sst = \nth{n T}$. It follows that the combination $n^2 \kappa_\sst T$ ---
which through eq.~\thermalresult\ governs neutrino scattering --- simply reduces 
to the particle density, $n$. This leads to a neutrino scattering rate, $\Gamma
\sim \sigma n \sim \GF^2 m E n$, which for solar neutrinos in the sun's centre 
($E \sim 1$ MeV and $n_c \sim 10^{26}$/cm${}^3$) scattering from 
nucleons ($m \sim 1$ GeV) is negligibly small: $\Gamma^{-1} \sim 10^{10} \, \hbox{km}$. 

\section{Macroscopic Density Fluctuations}

Our second application of the formalism of section 2 is to {\it macroscopic}
variations in the mean currents, $\Avge{J^\mu_a}$, that arise in eq.~\thvbarone.
We do so partly because this source of fluctuations has until recently been 
ignored in the literature. More importantly, this type of fluctuation can 
produce effects which are much larger than those which arise microscopically. 
There are many situations in electromagnetism for which macroscopic 
fluctuations can furnish the dominant medium-dependent effects. A familiar
example is furnished by the case of light propagating through a 
cloud. In this case the cloud is opaque because of density fluctuations 
on the scale of the water droplets which make up the cloud, rather than 
fluctuations on more microscopic scales.\foot\notquite{Of course, 
this analogy can be misleading if applied too literally to neutrino physics, since
the absence of multiple scattering precludes  neutrinos from `refracting' from a
large scale density fluctuation in the sun in the  same way that light refracts
through a water droplet.}  
Our purpose here, and in Section 4, is to analyze the
analogue of such fluctuations for neutrinos. 

Eq.~\thermalresult\ implies that thermal fluctuations have negligible effects 
for neutrinos passing through the sun, so our starting point is the 
mean hamiltonian, eq.~\thvbarone, which describes neutrino
propagation after averaging over microscopic matter fluctuations:
\eq
\label\thvbaroneagain
V_\sse = - \int d^3x \;  \; j^\mu_a(x) \; i \, 
\nubar_i \gamma_\mu \left( \Pl g^a_{ij}  
+ \Pr h^a_{ij} \right) \nu_j ,
\eeq
where we write $j^\mu_a(x) = \Avge{J^\mu_a(x)}$ for the mean current.

Now comes the main point. For any fixed current profile, $j^\mu_a(x)$, 
the propagation of any particular neutrino through the sun is perfectly well
described by pure-state evolution using the mean hamiltonian given in 
eq.~\thvbarone (or, equivalently, eq.~\thvbaroneagain). (This point is 
demonstrated in detail in Section 5.) It is, however, {\it not} in general 
true that successive neutrinos see the {\it same} profile, 
$j^\mu_a(x)$. On the contrary, successive
neutrinos arriving at a detector may have been produced at different places
within the sun and so can pass through entirely different density profiles 
while {\it en route} to the earth. Alternatively, the density profile itself 
can change in the interval between the passage through the same region 
of different neutrinos. As a result, the neutrino flux to which a detector 
is exposed can be thought to have been processed through a constantly
changing kaleidoscope of density profiles. 

We wish to adapt the formalism of Section 2 to describe the influence on
neutrinos of this eternally varying current profile. We do so by modelling
these density variations as being random in character. We therefore consider
passing neutrinos through an {\it ensemble} of density profiles --- whose
properties are elaborated below --- over which we must average to obtain
the neutrino signal as seen by a detector on earth. Taking advantage 
of the recursive nature of the formalism of Section 2, we may simply
take these formulae over in whole cloth, but with the mean hamiltonian 
of eq.~\thvbarone\ now interpreted as the microscopic hamiltonian,
and with the averages over sector $B = \Sce$ (`$\Sce$nsemble') 
now interpreted as ensemble averages.  

The results are immediate. To first order in $V_\sse$, the mean 
hamiltonian after the ensemble average now becomes: 
\eq
\label\thvbaroneens
\Vbar_1 = - \int d^3x \;  \; \Avgce{j^\mu_a(x)} \; i \, 
\nubar_i \gamma_\mu \left( \Pl g^a_{ij}  
+ \Pr h^a_{ij} \right) \nu_j .
\eeq
Similarly, at second order we find:
\eq
\label\thvbartwoens
\eqalign{
\Vbar_2 &= i  \int_{t'}^t d\tau \, d^3x \, d^3x' \; \nubar_i \gamma_\mu 
\left( \Pl g^a_{ij}  + \Pr h^a_{ij} \right) \nu_j \cr
& \qquad\qquad\qquad\qquad  \nubar_k' \gamma_\lambda \left( \Pl g^b_{kl}  + 
\Pr h^b_{kl} \right) \nu_l' \; \Avgce{\delta j^\mu_a \delta j^\lambda_b{}'} , \cr}
\eeq
and
\eq
\label\thdiffens
\eqalign{
\left({\partial \rho^d_\nu \over \partial t}\right)_2 &= - \int_{t'}^t d\tau 
\, d^3x \, d^3x' \; \Bigl\{ \nubar_i \gamma_\mu  \left( \Pl g^a_{ij}  
+  \Pr h^a_{ij} \right) \nu_j \cr
& \qquad\qquad  \rho_\nu \; \nubar_k' \gamma_\lambda   
\left( \Pl g^b_{kl} + \Pr h^b_{kl} \right) \nu_l' \;  \Avgce{
\delta j^\lambda_b{}' \delta j^\mu_a } + 
(t \leftrightarrow \tau) \Bigr\},  \cr}
\eeq
where $\delta j^\mu_a(x) \equiv j^\mu_a(x) - \Avgce{j^\mu_a(x)}$. 

What remains is to estimate the ensemble averages, 
$\Avgce{j^\mu_a(x)}$ and $\Avgce{\delta j^\lambda_a(x) \delta j^\mu_b(x')}$,
which appear in these expressions. A key difference between these 
averages and those considered previously is that we may no longer assume 
the currents to be delta-correlated, as in eq.~\jjdeltaapprox. 

\vfill\eject

\subsection{The Ensemble Properties}

The precise nature of these ensemble averages depends on the kinds of physics
which is responsible for the varying currents that successive neutrinos see. 
It is useful to have a systematic framework within which to couch our later
models of these fluctuations. This section outlines such a framework.

Suppose, then, that the currents $j^\mu_a(x)$ are expanded in terms of a
complete set of orthonormal functions, $\phi_\ssn(x)$, as follows:
\eq
\label\modeexpansion
j^\mu_a(x) = \jbar^\mu_a(x) \left[ 1 + \sum_N \Scc_{a\ssn} 
\, \phi_\ssn(x) \right] .
\eeq
We take the coefficients, $\Scc_{a\ssn}$, to be random variables 
having vanishing mean, which are uncorrelated for different modes:
\eq
\label\randomdefs
\Avgce{\Scc_{a\ssn}} = 0 \qquad \hbox{and} \qquad \Avgce{
\Scc_{a\ssn} \, \Scc_{b\ssm}} = C^2_{ab\ssn} \; 
\delta_{\ssn\ssm} .
\eeq
This implies the following for the density distributions
themselves:
\eq
\label\randomn
\eqalign{
\Avgce{j^\mu_a(x)} &= \jbar^\mu_a(x) , \cr
\Avgce{\delta j^\mu_a(x) \, \delta j^\nu_b(x')} &= \jbar^\mu_a(x) \, 
\jbar^\nu_b(x') \sum_N C^2_{ab\ssn} \, \phi_\ssn(x) \, 
\phi_{\ssn}(x'). \cr} 
\eeq
Notice that the completeness of the basis functions implies that
the currents, $j^\mu_a(x)$, become microscopically correlated, $\Avgce{
\delta j^\mu_a(x) \, \delta j^\nu_b(x')} \propto \delta( \bfx- \bfx')$, 
if $C^2_{ab\ssn}$ should be the same for all $N$. 

Clearly it cannot be true that the variables, $\Scc_{a\ssn}$, are uncorrelated
for all choices of basis functions. Different physical origins for the 
underlying randomness can lead to a different choice for the preferred, 
uncorrelated, basis. In what follows we use the following two models for the 
density fluctuations in the sun.

\topic{Locally-Varying Density Fluctuations}

As our first model of solar fluctuations we picture $j^\mu_a(x)$ to be
varying randomly from place to place. Motivated by the picture of
the solar medium consisting of turbulent regions of fluid we imagine 
dividing the sun into cells, labelled by the index $N$, whose volume  
we denote by $V_\ssn$. We permit this volume to vary slowly
as one moves around the sun. We then choose the basis functions to
be:
\eq
\label\randmodel
\phi_\ssn(x) = \left\{ \matrix{V_\ssn^{-\hf} & \hbox{if $\bfx$ lies within
cell $N$}, \cr 0 & \hbox{otherwise} . \cr } 
\right.
\eeq
With this choice we find the correlations:
\eq
\label\cellcorr
\Avgce{\delta j^\mu_a(x) \, \delta j^\nu_b(x')} = 
\jbar_a^\mu(x) \jbar_b^\nu(x')  \; 
\left\{ \matrix{ \epsilon^2_{ab\ssn}  
& \hbox{if $\bfx$ and $\bfx'$ both lie within
cell $N$}, \cr 0 & \hbox{otherwise} . \cr } 
\right.
\eeq
We introduce here the dimensionless quantity $\epsilon_{ab\ssn}$ by:
$C^2_{ab\ssn} \equiv \epsilon^2_{ab\ssn} \; V_\ssn$, to remove the. 
dependence on the cell volume, $V_\ssn$, which enters 
due to the normalization condition for the basis functions,
$\phi_\ssn(x)$. 

\ref\convectionref{See H.C. Spruit, {\it Mem. S.A. It.} {\bf }
(1996), astro-ph/9605020.}

How big might $\epsilon_{ab\ssn}$ and $V_\ssn$ reasonably be expected to be? For
solar applications the convective zone is known to contain density variations
on many scales \convectionref. Granules on the solar surface are 
$\sim 100$ km across. Giant convection cells are believed to have 
dimensions which are comparable to the depth of the convective zone itself: 
$\sim 2\times 10^5$ km.  Of more interest for neutrino propagation are 
the scales at the depths where neutrinos are produced, and where they resonate. 
Unfortunately, both of these regions lie within the radiative zone, where 
intuition based on the convective zone is unlikely to apply. Our analysis 
in subsequent sections of
how these fluctuations modify MSW oscillations indicates that $\epsilon^2 \ell
\gsim 100$ m is the range which is likely to have phenomenologically interesting
implications for neutrino oscillations, where $\ell$ is the length of a typical
cell in the direction of neutrino motion.  

This type of random model is very much in the spirit of refs.~\noisy\ and 
\noisytwo, for which the electron density is modelled as a random variable
that is delta-correlated in space. In fact, eq.~\cellcorr\ directly reduces
to the ensemble used in ref.~\noisy\ in the limit of negligible correlation
length (taken in ref.~\noisy\ to be $\ell = 10$ km), and when $\epsilon^2 \ell$ 
is taken to be constant. Ref.~\noisytwo\ makes a slightly different choice,
ensuring a small correlation length by continually adjusting $\ell$ to be 
a tenth of the neutrino matter oscillation length, as this varies throughout 
the sun. Differences between our results and those of ref.~\noisytwo\ do arise
for some regimes, which we believe to be due to this difference in treatment of 
the correlation length. 

Besides not assuming negligibly small correlation lengths, a more important 
difference between eq.~\cellcorr\ and  refs.~\noisy\ and \noisytwo\ is that 
we may take the fluctuations to vary differently as a function of position and time,
as may be appropriate for some kinds of solar physics. Our next example
presents an illustration of such a case. 

\topic{Helioseismic Waves: Oscillatory Normal Modes}

\ref\TC{See, for example, S. Turck-Chi\`eze and I. Lopes, \apj{408}{93}{347};\bk
S. Turck-Chi\`eze \etal, \prep{230}{93}{57}.}

Our second simple model of fluctuations in the sun is meant to model 
helioseismic $p$-waves. As a source of fluctuations through which neutrinos
propagate, helioseismic waves have the great advantage of actually
being known to exist. Furthermore, a fair amount is known about the
spectrum and amplitude of these waves \TC. 

\ref\moretocome{C.P. Burgess and D. Michaud, in preparation.}

In order to strip away as much extraneous detail as possible,
we start here with a simple wave within a rectangular geometry. 
(We report elsewhere on the results of more detailed modelling of 
neutrino interactions with helioseismic waves \moretocome.) That is,
we choose our neutrinos to be moving up the positive $z$ axis, through
a medium whose length in the $z$ direction is $L = 2R_\odot$.
A basis of modes which vanishes at the boundaries of this volume is
given by:
\eq
\label\wavebasis
\eqalign{
\phi_{+} &= N_{+}\; \cos\left( {2 \pi z \over \ell_+} \right) \,
\cos \left( {2 \pi t \over \tau_+} \right), \cr
\phi_{-} &= N_{-} \; \sin\left( {2 \pi z \over \ell_-} \right) \,
\cos \left( {2 \pi t \over \tau_-} \right), \cr}
\eeq
where we take the period and wavelength to be related in terms of the
speed of sound, $c_s$. 
To start with we take $c_s$ to be a constant, but we also present
some results with $c_s = c_s(z)$ chosen to more accurately mimic the
properties of the sun. For constant $c_s$, momentum is conserved in the
$z$ direction, and $\ell_\pm$ are determined by the boundary conditions to 
be a positive integer: $\ell_{+} = L/(n-\hf)$ and $\ell_{-} = L/n$. 

Finally, $N_{\pm} = N_\pm(r_\perp)$, denotes the dependence of the modes on the two other
directions, $r_\perp = (x,y)$, transverse to the direction of neutrino motion. For a
$z$-dependent speed of sound, $c_s(z)$, we take $N_{\pm}$ to also be plane waves,
as in eq.~\wavebasis, labelled by the conserved transverse momentum, $\bfk_\perp
= k_x \bfe_x + k_y \bfe_y$. For slowly varying wave amplitudes we then take
the wave number in the $z$-direction to be given by 
\label\dispreln
\eq
k_z^2 = {\omega^2 \over c^2_s(z)} - k_\perp^2,
\eeq
with $\omega = 2\pi / \tau_\pm$. In our subsequent numerical applications
we take the wave amplitude to vanish for those $z$ for which the resulting
$k_z$ is imaginary. 

There are several types of physics which might be expected to produce a 
normal modes with a randomly varying amplitude. First, even if the sun
were to be oscillating with a single mode, the
amplitude of this mode as seen by successive neutrinos would differ.
This is because, although any one neutrino sees an essentially static
density profile, (since the time scale for neutrinos to entirely escape the
sun is quite short --- $\ell \le L \sim$ several 
seconds --- compared to typical wave 
periods --- $\tau \sim$ several minutes) successive 
neutrinos can catch the wave
at differing points in its cycle. Neutrinos passing through at random 
times would therefore see a wave with a randomly varying amplitude. 

Of course, the real sun does not simply ring with constant amplitude
because various (poorly understood) processes permit energy to be
transferred into and out of the various normal modes. This leads to
additional randomness to the mode amplitudes, as seen from the neutrino's
perspective. 
\endtopic  

\section{Tracing Over the Neutrino Momenta}

We now fill in the neutrino part of the picture, and compute how the
fluctuations considered in the preceding Sections can 
enter into single-particle neutrino evolution.
To this end we again apply the results of Section 2 to trace over the
momentum/position part of the single-particle neutrino Hilbert space,
with the goal of deriving the explicit form for eq.~\mastereqn\ 
acting in neutrino flavour space, with which we can analyze 
resonant neutrino oscillations. 

Two issues must be borne in mind when applying the results of Section 2
in this way. First, since we wish to keep track of the second-order
effects, described above, due to matter fluctuations, we must use
the recursive form for the effective description which was given in
section 2.5. We therefore divide the neutrino sector itself into two 
subsectors, $A'$ and $B'$, with $B'$ consisting of the span of
all of the momentum states of the neutrino sector, whilst $A'$ 
comprises the sector labelled by neutrino spins and flavours. To avoid
confusion with our earlier choices for $A$ and $B$, we introduce the
new notation $A' = F$ (`flavour') and $B' = P$ (`position') for this part
of the analysis.

Second, since all practical measurements of neutrino flavour 
also involve a position measurement --- \ie\ neutrino $x$ is measured 
to be of flavour $y$ when it arrived at point $z$ --- we must also 
remember to adopt the formulation of Section 2.2, in which we perform
a partial measurement in sector $B' = P$.  This innocuous point has 
important implications for the form of the fluctuation terms in the
evolution equations for the reduced density matrix in flavour/spin space.

To proceed we must choose the density matrix which describes the initial neutrino
state. Assuming that the neutrino flavour/spin sector is initially uncorrelated 
with the neutrino momentum, $\varrho_\nu = \varrho_\ssf \otimes 
\varrho_\ssp$, we must choose an explicit form for $\varrho_\ssp$, 
in order to evolve the spin/flavour state, $\rho_\ssf$, forward in time.  
For a single neutrino, we would take this to be a pure, single-particle state,
$\varrho_\ssp =  \ket{\psi} \bra{\psi}$, describing an outgoing spherical wave
packet which starts at $t=t' = 0$ at the nucleus whose fusion produced the neutrino. The
spatial width, $\xi$, of this packet we imagine to be of negligible, microscopic,
dimensions. Since applications to neutrino oscillations involve observations of this
wave a very long way away from its centre, it suffices for our purposes to 
dispense with the spherical geometry and treat $\psi_{k,z_0}(p)$ as a plane wave packet, 
starting at $z=z_0$ at $t=0$, and travelling along the $z$ axis with average
momentum, $k\sim$ MeV:
\eq
\label\purestate
\braket{p}{\psi_{k,z_0}} = \psi_k(p) = \left[ {2  \xi \over \sqrt{2 \pi}} \right]^\hf 
\; e^{-\xi^2 (p_z - k)^2 - i p_z z_0} \; \delta(p_x) 
\; \delta(p_y) .
\eeq
$\ket{\psi_{k,z_0}}$ so defined is continuum normalized in the $x$ and $y$ directions. 
For solar neutrinos this pure state must be averaged over the 
initial distribution for producing such a neutrino within the sun. 

Next we must define the partial neutrino position measurement, which is
meant to express the fact that we know where neutrino measurements are 
performed: the Earth. We therefore choose observables of the form \uncorobs\
($\Sco = \Sco_\ssf \otimes \Sco_\ssp$) with 
\eq
\label\posmeasdef
\Sco_\ssp(t_m) \equiv \ket{\bfr;t_m} \bra{\bfr;t_m} ,
\eeq
corresponding to neutrino detection at the point,
$\bfr$, at a measurement time $t=t_m$. For a long-term exposure
to a constant flux (such as for solar neutrinos) we integrate over 
the appropriate range for $t_m$. 

\subsection{First-Order Effects}

With these choices we may now evaluate the quantity $\Vbar_1$. (Second order 
effects due to the averaging over neutrino momenta are explored in
the next section.) This will reproduce the usual MSW hamiltonian. The proper 
description of neutrino scattering, including the effects of matter 
fluctuations, is therefore found by tracing eqs.~\thvbaroneens, \thvbartwoens\ and
\thdiffens\ (or, for microscopic fluctuations, eqs.~\thvbarone, \thvbartwo\ and \thdiff)
over the neutrino momentum sector. 

This trace is straightforward to perform, subject to two important approximations. 

\topic{Negligible Neutrino Masses}
The first of these is the assumption, previously encountered in Section 3, that all
neutrino masses may be neglected when evaluating $\Vbar_1$. This is a good approximation
for the masses and mixings which are relevant for solar neutrino oscillations.  
(Of course, neutrino masses {\it do} play an important role once $\Vbar_1$ is used to
evolve neutrino states forward in time.)

\topic{Slowly-Varying Density Profile} 
Secondly, $j^\mu_a(x)$ is assumed to vary negligibly over distances comparable to the
packet width, $\xi$, and to the neutrino wavelength, $\lambda_\nu = 2\pi/k$. 
This approximation implies the only significant scattering from macroscopic
fluctuations is in the forward direction. For applications to solar 
neutrinos $j^\mu_a$ varies macroscopically while $\xi$ and
$\lambda_\nu$ are of atomic dimensions, so this last 
approximation also holds extremely
well. 
\endtopic

We find the following effective hamiltonian, to first order in $V$:  
\eq
\label\meanvone
\eqalign{
[\Vbar_1(\bfr,t,t_m)]_{i\lambda;j\sigma} &= - i  \int d^3x \, d^3q \; 
\psi_{k,z_0}^*(\bfr,t_m) \psi_{k,z_0}(\bfq) \; j^\mu_a(\bfx,t) \cr
& \qquad\qquad\qquad\qquad \bra{\bfr,\lambda,i;t_m} \; \nubar \gamma_\mu
(\Pl g^a + \Pr h^a) \nu \; \ket{\bfq,\sigma,j} \cr
&\approx \nbar_a(\bfr,t) \; \Bigl[ \Scm^a_{ij} \; \theta(-\sigma) 
- \Scm^{a*}_{ij} \; \theta(\sigma) \Bigr] \; \delta_{\lambda\sigma}. \cr}
\eeq
This result acts trivially on the spin labels, $\lambda$ and $\sigma$, involving
only the step function, $\theta(\sigma) = \hf \, [1 + \hbox{sign}
\,\sigma]$, which projects onto left-handed (LH: $\sigma = - \hf$) and right
handed (RH: $\sigma = + \hf$) states. We label the spin space using
the projection of the spin in the $z$ (or propagation) direction. For massive 
neutrinos this choice is made in the neutrino rest frame, while for massless
neutrinos it applies in any Lorentz frame. Since $\lambda = - \hf$ corresponds to a
left-handed state, we see that the within the Standard Model the states for which
$\lambda = + \hf$ are antineutrinos. 
For $N_\nu$ neutrino species the $N_\nu$-by-$N_\nu$ 
matrices $\Scm^a_{ij}$ represent the action of
$\Vbar_1$ on the flavour indices, $i$ and $j$. They are given explicitly in terms
of the coupling matrices by $\Scm^a_{ij} = g^a_{ij}  - h^a_{ji} = g^a_{ij} -
h^{a*}_{ij}$. 

Finally, the quantity $\nbar_a(\bfr,t)$ 
in this equation denotes the following:
\eq
\label\njdefs
\nbar_a(\bfr,t) = 
\jbar^0_a[r_\perp,z_0 + vt,t] - \jbar^z_a[r_\perp,z_0 + vt,t] ,
\eeq
where $r_\perp=(r_x,r_y)$ is the measurement position transverse to the
neutrino propagation direction, and $v$ denotes the speed, $v = k/E_k \approx 1$,
associated with the central momentum, $k$, of the wave packet. Recall that
$z_0$ denotes the point of origin of the neutrino, which is to be averaged
at the end of the calculation. 

Using eq.~\nbarests\
for the mean currents, $\jbar_a^\mu$, together with the nonrelativistic 
approximation, which permits the neglect of $\jbar^z_a$ --- certainly good for the
electrons and nucleons within the sun --- we see that eq.~\meanvone, is recognizable as
the standard MSW  starting point for analyzing resonant neutrino mixing in matter. With
this encouragement, we now proceed to compute the second-order contributions to
neutrino evolution. 

\subsection{Second-Order Contributions}

The second-order contributions come in two kinds, as can be seen from 
eqs.~\recmastereqn\ and \recptbnvtwo\ of Section 2.5. First and foremost, 
there are the matter
fluctuations --- \ie\ eqs.~\thvbartwoens\ and \thdiffens\ due to macroscopic 
fluctuations in the ensemble. But, \apriori, there could 
also be diffuse contributions due to the neutrino momentum average 
{\it itself}, as was first discussed in ref.~\Sawyer. 

In this section we apply the general treatment of Section 2 to both types of fluctuations.
For matter fluctuations we do so using the two ensemble models which were introduced
in Section 4.1. The effective hamiltonian we obtain in this way will turn out to
have interesting implications for MSW oscillations in the next section. 
By contrast, we find no phenomenologically interesting
effects for solar neutrinos due to fluctuations which arise due to
integrating out the neutrino momenta. Since this
conclusions differs somewhat from that of ref.~\Sawyer, we reproduce
his results in our formalism, and show why our conclusions differ. 

We start with the formalism
of Section 2, with the following three sectors: ($i$) $A' = F$ for neutrino
spins and flavour; ($ii$) $B' = P$ for neutrino momenta and position; and
($iii$) $B = \Sce$ (or $E$) for the matter degrees of freedom. Using the same 
approximations as were used to obtain eq.~\meanvone\ for $\Vbar_1$, 
we find the second-order contribution to ${\partial \rho_\ssf \over \partial t}$
to have the following form, regardless of the source of fluctuations:
\eq
\label\newvbartwo
\eqalign{
(\Vbar_2)_{i\lambda;j\sigma} &\approx  -i \Sca_{ab}(\bfr,t,t_m) \; 
\Bigl[ (\Scm^a \Scm^b )_{ij} \; \theta(-\sigma) + (\Scm^{a*} \Scm^{b*})_{ij} \; 
\theta(\sigma) \Bigr] \;  \delta_{\lambda\sigma}, \cr
\left( {\partial \rho^d_\ssf \over \partial t} \right)_{i\lambda;j\sigma} 
&\approx 2 \Sca_{ab}(\bfr,t,t_m) \Bigl[ (\Scm^a \, \rho_\ssf \, \Scm^b )_{ij} 
\; \theta(-\sigma) + (\Scm^{a*} \, \rho_\ssf \, \Scm^{b*})_{ij} \; 
\theta(\sigma) \Bigr] \; \delta_{\lambda\sigma} . \cr}
\eeq
(Recall here $\bfr$
represents the position where the neutrino is detected at time $t=t_m$.)
The two kinds of fluctuations discussed above differ only in their 
predictions for the key coefficient, $\Sca_{ab}(\bfr,t,t_m)$. We now give expressions
for this quantity for each of the two cases.

\topic{Fluctuations due to Tracing out Neutrino Momenta}

Before computing $\Sca_{ab}$ for matter fluctuations, we briefly pause to
discuss the fluctuations which arise on integrating out the
neutrino momentum sector. We do so partly to make explicit the contact with 
ref.~\Sawyer. We also do so partly because such fluctuations can arise
and may be important in some circumstances. We argue here why solar
neutrinos are unlikely to be one such case.

A direct application of the formulae of Section 2 to this type of second-order
contributions shows them to vanish, within the approximations outlined above. 
This is because we have chosen to measure the neutrino position arbitrarily
accurately, at precisely one point, $(\bfr,t_m)$. It is therefore instructive 
to integrate the observable, $\Sco_\ssp$ of eq.~\posmeasdef,
over a finite detector volume, $\Scd$. That is, we now replace eq.~\posmeasdef\
by:
\eq
\label\detectorvolume
\Sco_\ssp = \int_\Scd d^3r \; \ket{\bfr,t_m} \bra{\bfr,t_m}.
\eeq

With these choices, formulae \recmastereqn\ and \recptbnvtwo, when applied to
the neutrino-sector trace, give eqs.~\newvbartwo, with 
\eq
\label\adefneutrino
\Sca_{ab}(\bfr,t,t_m) \approx 
\int_{t'}^t d\tau \; \Avg{\delta \nbar_a[r_\perp,z_0+vt,t] \; 
\delta \nbar_b[r_\perp,z_0+v\tau,\tau]}_\ssp. 
\eeq
Here $\nbar_a$ is as defined in eq.~\njdefs, and the average is over
the detector volume transverse to the neutrino momentum. For any
quantity, $A(\bfr,t)$, this average is defined by:
\eq
\label\transvav
\Avg{A(\bfr,t)}_\ssp \equiv {1 \over \Scd_\perp} \int_{\Scd_\perp} d^2r_\perp
\; A(r_\perp,r_z,t) ,
\eeq
where $\Scd_\perp$ denotes the area which the detector
presents transverse to the neutrino beam. As usual, $\delta \nbar_a$, 
denotes the deviation of $\nbar_a$ from this
transverse mean: $\delta \nbar_a[\bfr,t] \equiv \nbar_a [\bfr,t] -
\Avg{\nbar_a[\bfr,t]}_\ssp$.  The key point here is that
this deviation vanishes, $\delta \nbar_a = 0$, in the limit that the
transverse detector size, $\Scd_\perp$, is much smaller than the scales
over which $\nbar_a$ varies appreciably. 

This result makes sense physically. In the absence of matter fluctuations,
neutrino evolution in the presence of a fixed density profile can be computed
as an exercise in scattering from a fixed potential. Scattering only arises from
variations, $\delta n$, in the density from its spatial average. The main point is 
that the interference term between the scattered and initial waves in this problem 
is proportional to $\int_{\Scd_\perp} d^2r_\perp \;\delta n(r_\perp)$, and so
vanishes only if the transverse area of the detector is sufficiently large on the
scales over which $\delta n$ varies. It follows that the scattering is 
incoherent only for such large detectors.

The relation of this result with that of ref.~\Sawyer\ is now clear. In this reference,
the reduced density matrix for neutrino
flavours is defined by completely tracing over all neutrino momenta, without taking
into account the position measurement, eq.~\posmeasdef. This is equivalent to
taking the detector volume to fill all space, and our eqs.~\adefneutrino\ and
\transvav\ indeed reduce to ref.~\Sawyer's in this limit. But working in the limit
of an extremely large detector misses the important suppression of these effects
by the detector size. 

Since we find in later chapters that significant neutrino effects require fluctuations
on scales of hundreds of metres and up, we are led to conclude that this kind of 
neutrino incoherence likely plays no role for solar neutrinos.  

\topic{Matter Fluctuations}

For matter fluctuations, the second-order contribution to the neutrino evolution
equation is given by eq.~\newvbartwo, with  
\eq
\label\adefmatter
\Sca_{ab}(\bfr,t,t_m) = 
\int_{t'}^t d\tau \; \Avgce{ \delta n_a[r_\perp,z_0+vt,t] \; 
\delta n_b[r_\perp,z_0+v\tau,\tau]}. 
\eeq
As before $n_a$ denotes the difference $j^0_a - j^z_a$ (or simply
the density, $j^0_a$, in the nonrelativistic limit), 
while $\delta n_a$ is the difference between
$n_a$, and its ensemble average, $\nbar_a = \Avgce{n_a}$. 

Eq.~\adefmatter\ may be explicitly computed within the two models of fluctuations
which were introduced in Section 4. For the case of locally-varying density
fluctuations, eq.~\randmodel, we have:
\eq
\label\lvdfmodfora
\eqalign{
\Sca_{ab}(\bfr,t,t_m) &= \epsilon_{ab\ssn} ^2  \; \nbar_a[
r_\perp,z_0+vt,t] \int_{\hbox{cell $N$}} \; \nbar_b[
r_\perp,z_0+v\tau,\tau] \; d\tau \cr
&\approx \epsilon_{ab\ssn}^2 \, \ell_\ssn \; \nbar_a[
r_\perp,z_0+vt,t] \; \nbar_b[ r_\perp,z_0+vt,t] \cr 
& \qquad\qquad\qquad \hbox{(locally-varying density 
fluctuations)} .\cr}
\eeq 
Here $\ell_\ssn$ is the length of cell $N$ along the neutrino line of
flight, and $N$ labels the specific cell which contains the point 
$(r_\perp,z_0+vt,t)$. The approximate equality in the second line 
is derived under the assumption that the mean
current, $\nbar_a$, does not vary appreciably over the size of this cell.
For the approximate exponential density profile \SNTheory: 
\eq
\label\expform
\nbar_a(z) = (n_a)_c \; e^{-z/h} ,
\eeq
which we use for electrons in the sun, the neglect of the variation of $\nbar_a$
over a cell requires $\ell \ll h = R_\odot/10.5 = 6.6 \times 10^4$ km.
(Notice this is a much weaker condition than requiring $\ell$ to
be much smaller than neutrino propagation scales.) 
For electrons, the central density 
we use is $(n_a)_c = 1.5 \times 10^{26}/$cm${}^{3}$. 

Similarly, $\Sca_{ab}(\bfr,t,t_m)$ may also be evaluated
for the case of an oscillatory density profile. 
For solar applications, since $\tau_\pm$ is of order several minutes and
(the light-travel time across) $\ell_\pm \lsim R_\odot$ is of order a few seconds,  
it is sufficient to neglect powers of $\ell_\pm/\tau_\pm$ and $R_\odot/\tau_\pm$.
Again using the exponential profile, eq.~\expform, for $\nbar_z(x)$,
we find:
\eq
\label\oscresult
\eqalign{
\Sca_{ab}^\pm(r_\perp,t,t_m) &= (n_a)_c \, (n_b)_c \, \epsilon_\pm^2(r_\perp) 
 \; F_\pm[z_0+vt,t,t'], \cr
& \qquad\qquad\qquad \hbox{(oscillatory density 
fluctuations)} \cr}
\eeq
where $\epsilon_\pm^2 \equiv \Scc_\pm^2 N^2_\pm$ defines the dimensionless
size of the fluctuation, and the function $F$ is
defined by 
\eq 
\label\Fpmdef
F_\pm(z_0+vt,t,t') \equiv e^{-(z_0+vt)/h} f_\pm(z_0+vt) 
\int_{z_0+t'}^{z_0+t} dx \; e^{- x/h} \, f_\pm(x)
\eeq
with $f_-(z) = \sin\left( {2 \pi z \over \ell_-} \right)$ and $f_+(z) =
\cos\left( {2 \pi z \over \ell_+} \right)$. The integral is elementary and
is given (writing $a = 1/h$ and $b = 2\pi/\ell$) by:
\eq
\label\oscints
\eqalign{
F_-(z_0,t-t') &= { e^{-a(z_0+t)} \, \sin b(z_0+t) \over a^2 + b^2} \; 
\Bigl\{ e^{-a(z_0+t')} \; \Bigl[ b \cos b(z_0+t') + a \; \sin b(z_0+t') \Bigr] \cr
& \qquad\qquad\qquad\qquad\qquad\qquad\qquad
\qquad\qquad - (t' \to t) \Bigr\} , \cr  
F_+(z_0+t,t,t') &= {e^{-a(z_0+t)} \, \cos b(z_0+t) \over a^2 + b^2} \;
\Bigl\{ e^{-a(z_0+t')} \; \Bigl[ a \cos b(z_0+t') + b \; \sin b(z_0+t') \Bigr] \cr
& \qquad\qquad\qquad\qquad\qquad\qquad\qquad\qquad\qquad
- (t' \to t) \Bigr\} . \cr  }
\eeq
\endtopic

To summarize, fluctuations can indeed influence the neutrino evolution. Their
effects are quantified by equations \newvbartwo\ and \adefmatter, which are
the main results of this section. Their
implications for resonant MSW oscillations can be sizable, as is now
explored in more detail. 

\section{Applications to MSW Oscillations}

In this section we evolve the neutrino density matrix to second order in $\GF$. 
With solar neutrinos in mind we follow the usual practice and suppose the initial 
neutrino spin to be purely  left-handed, and focus on the evolution in flavour 
space. The plan is to use eqs.~\meanvone\ and \newvbartwo\ to evaluate the
right-hand-side of eqs.~\mastereqn\ and \drhodiff, and then to integrate the result to
determine the electron-neutrino survival probability, $P_e(t) = \rho_{ee}(t)$. 

\vfill\eject

\subsection{The Evolution Equations}

For simplicity we specialize also to the case of two neutrino flavours,
whose electroweak eigenstates we denote $e$ and $\mu$, although we
might equally well imagine mixing the $e$-type and $\tau$-type 
neutrinos. Eq.~\markoveqn\ then takes the following form:
\eq
\label\explme
\eqalign{
{\partial \rho \over \partial t} &= -i \Bigl[ V_0 + \Vbar_1, \rho \Bigr] - 
2 \, \GF^2 \sum_{a,b=e,n} \Sca_{ab} \; (g^a g^b \rho + 
\rho g^a g^b - 2 g^a \rho g^b), \cr
&= -i \Bigl[ V_0 + \Vbar_1, \rho \Bigr] - 2 \, \GF^2 \Sca_{ee} \,
\Bigl[(g^e)^2 \rho + \rho (g^e)^2 
- 2 g^e \rho g^e \Bigr], \cr}
\eeq
where $\rho = \rho_\ssf$ is the neutrino-sector density matrix in flavour space, and
\eq
\label\explmedefs
\eqalign{ 
V_0 & \equiv \Bigl[ k^2 + m^\dagger m \Bigr]^{1/2} \approx k + 
{m^\dagger m \over 2k} + \cdots, \cr
\Vbar_1(t) &\equiv \sqrt2 \, \GF \Bigl[ g^e \; \nbar_e(t) +
g^n \; \nbar_n(t)  \Bigr] .\cr}
\eeq
Here $m$, $g^e$ and $g^n$ are $2 \times 2$ matrices which represent
the left-handed-neutrino mass matrix, and the neutrino charged- and
neutral-current coupling matrices. In an electroweak basis these are
given explicitly by:
\eq
\label\thematrices
m = \pmatrix{ m_{ee} & m_{e\mu} \cr m_{e\mu} & m_{\mu\mu} \cr},
\qquad g^e = \pmatrix{1&0\cr 0&0\cr}, \qquad \hbox{and} \qquad
g^n = \pmatrix{-\, \hf & 0 \cr 0 & - \,  \hf \cr}. 
\eeq
In general $m$ may  be a generic symmetric complex matrix, although in 
the absence of CP violation it may be chosen to be real. $\nbar_e(t)$ and 
$\nbar_n(t)$ are the spatially-averaged electron and neutron currents, 
as defined in eq.~\njdefs. 

The second-order contribution to eq.~\explme\ is given by the three terms
proportional to $\Sca_{ab}(t)$, which is defined for $a,b = e,n$ by eq.~\adefmatter. 
The terms of the form $g^2 \rho$ and $\rho g^2$ are the contributions due to
$\Vbar_2$, while the $g \rho g$ term comes from ${\partial \rho^d_\ssf \over \partial t}$.
Notice that because $g^n$ is proportional to the unit matrix, 
all but the term involving $\Sca_{ee}$ --- which we henceforth denote 
simply by $\Sca$ --- give zero in the sum in the first of eqs.~\explme.
As a result, it is only fluctuations in the electron density profile which are
relevant for neutrino evolution in the sun. 

In order to integrate this equation it is useful to expand all matrices
in terms of the unit and Pauli matrices, $\{I,\vec\tau\}$. We have
\eq
\label\pauliexpn
\eqalign{
\rho = \rho_0 + \vec\rho \cdot \tau  \qquad &\hbox{and} \qquad
V_0 = M_0 + \vec M \cdot \tau, \cr
g^e = \hf \; (1 + \tau_3), \qquad &\hbox{and} \qquad g^n = - \,  \hf , \cr}
\eeq
with 
\eq
\label\rhocomps
\rho_{ee} = \rho_0 + \rho_3, \qquad \rho_{\mu\mu} = \rho_0 - \rho_3, 
\qquad \rho_{e\mu} = \rho_1 -i \rho_2, 
\eeq
and
\eq
\label\Mcomps
\eqalign{
M_0 = k + {|m_{e\mu}|^2 + \hf \; (|m_{ee}|^2 + |m_{\mu\mu}|^2) \over 2k} , &\qquad
M_1 =  {\Re (m_{ee}^*m_{e\mu} + m_{e\mu}^* m_{\mu\mu}) \over 2 k},  \cr
M_2 = - \,  {\Im (m_{ee}^*m_{e\mu} + m_{e\mu}^* m_{\mu\mu}) \over 2 k},  &\qquad
M_3 = {|m_{ee}|^2 - |m_{\mu\mu}|^2 \over 4 k} . \cr}
\eeq
Of these components, $M_0$ plays no role in the evolution of $\rho$ since
it drops out of the right-hand-side of eq.~\explme. Similarly, since the 
trace of the right-hand side of eq.~\explme\ vanishes identically, it follows that the
coefficient $\rho_0$ is independent of time: ${\partial \rho_0 \over \partial t} = 0$.  

With these definitions the evolution equation, \explme, for the remaining 
three components of $\rho$ may be written:\foot\others{If $\ss 
\Sca(t)$ is assumed to be a constant times $\ss \nbar^2(t)$, and in the 
absence of CP violation, then this equation agrees with that 
used in ref.~\noisy. It also agrees
with ref.~\noisytwo\ if the correlation length is adjusted as
explained in Section 4.1.}
\eq
\label\finalform
{\partial \over \partial t} \pmatrix{\rho_1 \cr \rho_2 \cr \rho_3 \cr}
= \pmatrix{- 2 a & -2(M_3 + b) & 2 M_2 \cr 2 (M_3 + b) & -2 a & -2 M_1 \cr
-2 M_2 & 2 M_1 & 0 \cr} \; \pmatrix{\rho_1 \cr \rho_2 \cr \rho_3 \cr}
\equiv \Sch \pmatrix{\rho_1 \cr \rho_2 \cr \rho_3 \cr} ,
\eeq
where 
\eq
\label\abdefs
a(t) \equiv \GF^2 \Sca(t) \qquad \hbox{ and} \qquad b(t) \equiv {\GF \nbar_e(t) 
\over  \sqrt2} .
\eeq
This is the form for the evolution which are integrated in subsequent sections. 

For the purposes of exploring the implications of electron density fluctuations
it suffices to restrict our attention to CP-conserving neutrino physics, for which
the components of $V_0$ simplify somewhat because the
neutrino mass matrix, $m_{ij}$, may be chosen to be real. For this case
$M_1$ through $M_3$ can be expressed in terms of the heavy  and light
neutrino mass eigenvalues, $m_h$ and $m_l$, by:
\eq
\label\Mcompstwo
M_1 = {\delta m^2 \sin 2 \theta_\ssv \over 4 k}, \qquad
M_2 = 0,  \qquad 
M_3 =  - \, {\delta m^2 \cos 2 \theta_\ssv \over 4 k} ,
\eeq
where $\delta m^2 = m_h^2 - m_\ell^2$ and $\theta_\ssv$ is the vacuum
mixing angle, for which 
\eq
\label\thetadef
\nu_\ell = \nu_e \; \cos\theta_\ssv - \nu_\mu \; \sin\theta_\ssv, 
\qquad 
\nu_h = \nu_e \; \sin\theta_\ssv + \nu_\mu \; \cos\theta_\ssv .
\eeq 

We now turn to integrating eq.~\finalform. Before turning to a numerical
solution using an exponentially falling electron density profile \SNTheory,
we first pause for the instructive exercise of solving this equation in the
adiabatic and Parke limits. 

\subsection{Adiabatic Evolution and the Generalized Parke Formula}

If $a$ and $b$ are slowly varying, then it is straightforward to analytically integrate
eq.~\finalform\ simply by performing a ($t$-dependent) rotation which diagonalizes
the matrix $\Sch$. Once these adiabatic solutions are obtained, then Parke's more general
expression may then be derived by starting with these states as bases and 
computing the transition probability as the neutrinos pass through the resonance.

The adiabatic result is:
\eq
\label\genintegral
\pmatrix{\rho_1 \cr \rho_2 \cr \rho_3 \cr}(t)  = R(t) \pmatrix{e^{\int_{t'}^t
\lambda_1(x) dx}
& & \cr & e^{\int_{t'}^t
\lambda_2(x) dx} & \cr && e^{\int_{t'}^t
\lambda_3(x) dx} \cr} \; R^\dagger(t') \;
\pmatrix{\rho_1 \cr \rho_2 \cr \rho_3 \cr}_{t = t'}, 
\eeq
where $\lambda_i(t)$ are the three time-dependent
eigenvalues of $\Sch$, and $R(t)$ is the matrix
for which $R^\dagger(t)  \Sch(t)\ R(t) = \diag{\lambda_1(t), \lambda_2(t),
 \lambda_3(t)}$. 

Keeping in mind that $a$ arises at second order in $\GF$ and so is smaller
than all of the other elements, we may solve for the $\lambda_i$ and $R$
perturbatively in $a$. The resulting eigenvalues are, to linear order in $a$:
\eq
\label\evals
\lambda_0 = - 2\gamma_0 , \qquad \lambda_\pm = \pm 2i \kappa - 2\gamma,
\eeq
with 
\eq
\label\evalsstuff
\kappa = \sqrt{M_1^2 + (M_3 + b)^2}, \qquad \gamma_0 = {a M_1^2 \over
M_1^2 + (M_3 + b)^2} , \qquad \gamma = {a [M_1^2 + 2(M_3 + b)^2 ]\over
2[M_1^2 + (M_3 + b)^2]}.
\eeq
To the lowest (zeroeth) order in $a$ the corresponding matrix, $R$, is:
\eq
\label\evecs
R = \pmatrix{{M_1 \over \kappa} & - \, {M_3 + b \over \sqrt2 \kappa} & 
- \, {M_3 + b \over \sqrt2 \kappa} \cr 0 & {i \over \sqrt2} & - \,  {i \over \sqrt2}  \cr
{M_3 + b \over \kappa} & {M_1 \over \sqrt2 \kappa} & {M_1 \over \sqrt2 \kappa} \cr}
 = \pmatrix{\sin 2\theta_m & \nth{\sqrt2} \, \cos 2\theta_m & 
\nth{\sqrt2} \, \cos 2\theta_m  \cr 0 & {i \over \sqrt2} & - \,  {i \over \sqrt2}  \cr
- \cos 2\theta_m  & \nth{\sqrt2} \, \sin 2\theta_m  & \nth{\sqrt2} \, 
\sin 2\theta_m  \cr}, 
\eeq
where the last equality defines the usual matter mixing angle, $\theta_m$. 

Using eqs.~\evals, \evalsstuff\ and \evecs\ in eq.~\genintegral\ then gives the
adiabatic prediction for the electron-neutrino 
survival probability, given the initial
condition $\rho(t') = \diag{1,0}$:
\eq
\label\peepred
\eqalign{
P_e(t) &\equiv \rho_{ee}(t) = \hf \; \left\{ 1 + e^{- 2 \int_{t'}^t 
\gamma_0(x) dx} \, \cos 2\theta_m(t') \, \cos 2\theta_m(t) \right. \cr
& \left. \qquad \qquad + e^{-2 \int_{t'}^t \gamma(x) dx} \, \sin 2\theta_m(t) \, 
\sin 2\theta_m(t') \cos \left[ 2 \int_{t'}^t \kappa(x) dx \right]  
\right\} . \cr}
\eeq
The oscillatory term, on the second line of eq.~\peepred, averages to zero once we sum
over a long enough measurement interval. 

\ref\Parke{S.J. Parke, \prl{57}{86}{1275}.}

For nonadiabatic evolution it is straightforward to use these adiabatic
results to derive a generalization of Parke's formula. After averaging over
the production and detection times we find:
\eq
\label\Parkeform
P_e(t) = \hf + \left( \hf - P_\ssj \right) 
 e^{- 2 \int_{t'}^t \gamma_0(x) dx} \, \cos 2\theta_m(t') 
\, \cos 2\theta_m(t) ,
\eeq
in which $P_\ssj = \exp\left[ - \, {\pi \over 2} \; \left( {\sin^2 2 \theta_\ssv
\over \cos 2 \theta_\ssv} \right) \left( {\delta m^2 \, h \over 2 k}
\right) \right] $ is the `jump' probability as one passes through the
resonance point. 

There are many reasons to become emotionally involved with eq.~\Parkeform:

\topic{1}
In the absence of fluctuations, $a \to 0$, eqs.~\peepred\ and \Parkeform\ 
reduce to the standard results for matter oscillations. 

\topic{2}
When $a$ is small, but not zero, its dominant influence is to {\it damp} the 
neutrino oscillations, by introducing an imaginary part to the masses of the
mass eigenstates in matter. Such an imaginary contribution might have been expected
given the antihermitian form found in eq.~\newvbartwo\ for $\Vbar_2$. 

The resulting damped oscillations are similar to what arises when 
neutrinos decay, but with an important
difference. The difference is the appearance 
of the term ${\partial \rho^d_\ssf \over \partial t}$ in the
evolution equation, \mastereqn. As a result, there is no
net loss of probability in the neutrino sector: $\Tr \rho_\ssf(t) \equiv 1$ for
all $t$. For neutrinos the damping is a reflection of the conversion of
the incoming neutrino from a pure to a mixed state, due to the decoherence
introduced by the fluctuations in the matter through which it passes. 

\topic{3}
The relative size of the fluctuation parameter, $a = \GF^2 \Sca$, in comparison
with the usual MSW effective hamiltonian, $b = \GF \nbar/\sqrt2$, is given 
(for locally-varying fluctuations) by
\eq
\label\sizeparam
{a \over b} \sim \GF \nbar \, \epsilon^2 \ell \sim 
\left( {\epsilon^2 \ell \over 100 \, \hbox{km}} \right) \; \left( {\nbar 
\over 10^{26} \, \hbox{cm}{}^{-3}} \right) ,
\eeq
which gives a rough indication how large fluctuations must be in order to contribute
\break significantly to neutrino evolution. 

Important effects arise even for smaller $\epsilon^2 \ell$, however, in the case of
resonant oscillations. This is because even a small damping term can ruin the quality
of the resonance. This is borne out by our numerical integrations, which show that the
first modifications arise precisely at the resonant point. It can also be seen
analytically from eq.~\Parkeform, as we now argue.

\topic{4}
The strength of the decohering effect is completely parameterized by the integral
over $\gamma_0$ which appears in eq.~\Parkeform, after the oscillatory
contributions are averaged out. This integral is elementary (for an exponentially
falling electron density profile) when $\Sca(x)$ is a constant
times $n_e^2(x)$, as is the case for the locally-varying 
density-fluctuation scenario,
discussed earlier. In this case we have $\Sca = \epsilon^2 \ell \, n_e^2$ and so
\eq
\label\dampint
\eqalign{
\Scr &\equiv \exp\left\{ -2 \int_{t'}^{t_{\rm ex}} \gamma_0(x) \; dx \right\} 
= \exp\left\{ -2h \int_{0}^{n'} \gamma_0(n) \; {dn \over n} \right\} , \cr
&= \exp\left\{ - \, \left({\epsilon^2 \, \ell \, h \, (\delta m^2)^2 \sin^2
2\theta_\ssv \over 4 k^2} \right)
\left[ \ln\left| {\sin 2 \theta_\ssv \over \sin 2\theta_m'} 
\right| \right. \right. \cr
& \qquad \qquad  + \left. \left. \cot 2\theta_\ssv \Bigl( \arctan \cot 2 \theta_\ssv -
\arctan \cot 2 \theta_m' \Bigr) \right] \right\}, \cr
& \approx  \exp\left\{ - \,\left( {\pi \epsilon^2 \, \ell \, h \, (\delta m^2)^2 \sin
2\theta_\ssv \over 4 k^2} \right) \Bigl( 1 + O ( \sin 2 \theta_\ssv )
\Bigr) \right\}, \cr
& \approx \exp\left\{ - \, \left({\epsilon^2 \, \ell \over 7 \, {\rm km}} \right)
\, \left({\delta m^2 / k \over 1 \eV^2/\MeV} \right)^2 \, 
\left({\sin 2 \theta_\ssv  \over 0.1} \right) \right\} .\cr} 
\eeq
Here $t_{\rm ex}$ is the time the neutrino exits the sun, and so after which 
$n_e(t)$ and $\gamma_0(t)$ both vanish. 
$\theta_m' = \theta_m(t')$ and $n' = n_e(t')$
are the matter mixing angle and electron density 
at the production point, deep within the solar interior. 

Eq.~\dampint\ shows that significant damping is possible for solar neutrinos when
$\epsilon^2 \ell \gsim$ a few kilometers. 

\topic{5}
For the smallest amplitude fluctuations, the first place where the
damping becomes noticeable is for resonant, adiabatic oscillations (see
Figures 1 and 2). That is, for $P_\ssj \approx 0$ and $\cos 2\theta_m' \approx
-1$, and taking $\theta_m(t)= \theta_\ssv$, eq.~\Parkeform\ becomes:
\eq
\label\adiabaticosclim
P_e \approx \hf \left[ 1 -  e^{- 2 \int_{t'}^t \gamma_0(x) dx} \, 
\cos 2\theta_\ssv \right] . 
\eeq
For $a=0$ (no fluctuations) this gives the usual suppression: $P_e(t)
\approx \sin^2\theta_\ssv$. For small damping this suppression is 
weakened. 

Since the success of MSW oscillations in explaining the solar-neutrino
data uses this suppression to virtually remove the ${}^7$Be neutrino line,
a good measurement of the strength of this flux promises to give information
about the strength of electron density fluctuations deep within the solar
interior.  

\topic{6}
For sufficiently large times, $\int_{t'}^t \gamma_0(x) dx \gg 1$, eq.~\Parkeform\ has 
the universal prediction: $P_e \to \hf$. (This limit is also seen in our
numerical integrations, as well as in those of ref.~\noisy. We believe its absence
in ref.~\noisytwo\ is due to the use there of a correlation length which 
follows the neutrino matter oscillation length.) 
This suggests a new solution for the solar
neutrino problem: an approximately energy-independent suppression of the solar neutrino
flux by a factor of 2 due to solar fluctuations, even for small neutrino mixing angles
in vacuum. This type of solution will be disfavored if improvements in the data should
confirm the present indications for an energy-dependent neutrino suppression (\ie\
for which $p-p$ neutrinos are untouched while ${}^7Be$ neutrinos are 
essentially removed).

\topic{7}
Resonance is defined by the condition $M_3 + b = 0$. Interestingly, 
although this resonance condition minimizes $\gamma$,
it actually {\it maximizes} $\gamma_0$. This is clearest when eq.~\evalsstuff\ is
written: $\gamma_0(t) = a(t) \sin^2 2\theta_m(t)$. 
The sharper the resonance, the sharper the peak
there in $\gamma_0$. In the limit of a sudden resonance we therefore expect $\Scr$, of
eq.~\dampint, to be controlled by the value of $\Sca$ evaluated at the resonance point.
In this limit we expect the fluctuation size at the resonance point to be what
determines the size of the damping contribution. 
\endtopic

\subsection{Numerical Results}

We have numerically integrated eq.~\finalform\ to determine the electron-neutrino
survival probability, $P_e(t) = \rho_{ee}(t)$, using $\Sca(t)$ as predicted by both the
locally-varying density-fluctuation and the oscillating-mode model of density
fluctuations. For the oscillating-mode case we have examined both the case of constant
sound speed, and a speed which varies as a function of $z$.\foot\sunsoundspeed{ 
We use the form $\ss c(z)  = c_1 + c_2 \left( {z\over L}\right) + c_3 
\left( {z\over L}\right)^2 + c_4 \left( {z\over L}\right)^3 + c_5
\left( {z\over L}\right)^4$, in units of the speed of light, with coefficients $\ss c_1 =
0.00170$,  $\ss c_2 = 0.000581$, $\ss c_3 = - 0.0106$, $\ss c_4 = 0.0177$ and $\ss c_5 = -
0.00929$, as found by fitting to the profile given in ref.~\TC.} We use the exponential
density profile, eq.~\expform. 

The present section is devoted to presenting the results of these calculations. 
We have compared these numerical results with the analytical result, eq.~\Parkeform,
which we find works extremely well for all of the parameters of interest.

\figure\figone{The electron neutrino survival probability as a function of
$E/\delta m^2$ assuming the locally-varying density fluctuations model,
as described in the text. Each curve represents a different value for the parameter
combination $\epsilon^2 \ell$, where $\epsilon = \delta n/n$ is the fractional
amplitude of the fluctuation, and $\ell$ is the fluctuation's correlation
length in the direction of neutrino motion.}

\figure\figtwo{The electron neutrino survival probability as a function of
$E/\delta m^2$ assuming the oscillatory density fluctuations model, which is
a crude model of a helioseismic $p$-wave.  
Each curve represents a different value for the fractional amplitude. 
$\epsilon = \delta n/n$. The figure assumes a wave period of 30 minutes,
and uses the speed of sound as a function of depth given in ref.~\TC.
The transverse wavenumber, $k_\perp$, is chosen as small as possible, 
corresponding to a wave which penetrates as deeply as possible into the
solar interior.}

We present our results through Figures 1 and 2. These present plots of the
survival probability, $P_e(t)$, as a function of the neutrino energy normalized to
its squared mass: $E/\delta m^2$. We use $\sin 2 \theta_\ssv = 0.1$. In Figure 1 we use
the locally-varying density fluctuation model, with successive curves representing
different values for the fluctuation strength, $\epsilon^2 \ell$.  As expected based on
eq.~\Parkeform, sizable effects first appear by deteriorating the quality of the
suppression at resonance. The first deviations appear when $\epsilon^2 \ell = 0.01$
km, until the resonance is completely destroyed for $\epsilon^2 \ell = 100$ km. 
For small $E/\delta m^2$ and $\epsilon^2 \ell < 10$ km, the survival probability
approaches its asymptotic MSW form $P_e \approx \cos^2 2\theta_\ssv \approx 1$. For
$\epsilon^2 \ell$ larger than this value, $P_e$ instead tends to the limit $\hf$ for
small $E/\delta m^2$. 

Figure 2 presents a similar plot for an oscillatory fluctuation, using the
position-dependent speed of sound. The wave period is chosen to be 30 minutes,
and its transverse momentum, $\bfk_\perp$, is chosen to be as large as is
possible: $k_\perp \sim 2 \pi/L$, for $L \sim R_\odot$. The wavenumber,
$k_z$, in the direction of neutrino motion is fixed as a
function of $z$ using the dispersion relation, eq.~\dispreln, with a realistic
$z$-dependent speed of sound. As was described in section (4.1),
the wave's amplitude is set to zero when $k_z$ so determined is imaginary,
corresponding to a damped wave. Our plots are made using an `odd' mode, $\phi_-(x)$. 

The figure displays the resulting survival probability as a function of
$E/\delta m^2$ for various amplitudes, $\epsilon$. Sizable deviations start for
$\epsilon \gsim 1\%$. The size of this deviation roughly agrees with what would be
estimated using the locally-varying density fluctuation results of Fig.~1, with the
correlation length, $\ell$ taken as the wavelength at the resonant point.
This leads us to expect a larger effect in a more accurate simulation using 
$g$ waves than for $p$-waves, since these have larger amplitudes and wavelengths in
the resonance region. More realistic simulations to investigate these issues are under way
\moretocome. 

\section{Magnetic-Moment Couplings}

This section briefly applies the formalism of Section 2 to neutrinos which interact with
magnetic fields through a magnetic-moment interaction:
\eq
\label\magmomintdef
\eqalign{
\Scl &= {\mu_{ij} \over 4} \; [ \nubar_i \gamma^{\mu\nu} \Pl \nu_j ] \; F_{\mu\nu}
+ {\rm c.c.} \cr
&= {1 \over 4} \; \Bigl[ \nubar_i \gamma^{\mu\nu} ( \Re\mu_{ij} + i \gamma_5 \,
\Im\mu_{ij} ) \nu_j \Bigr] \; F_{\mu\nu} . \cr}
\eeq
Since the $\nu_i$ are majorana, the matrix $\mu_{ij}$ is (complex) antisymmetric.
Fluctuating magnetic fields have been considered in ref.~\noisy, in the limit of
negligibly short correlation length.

With this choice, the first-order contribution to the effective hamiltonian governing
neutrino evolution in spin and flavour space as it moves along the $z$ axis is:
\eq
\label\mmvbarone
(\Vbar_1)_{i\lambda;j\sigma} = \ol\Scf_{ij} \; \hbox{sign}(\sigma) \,
\delta_{-\lambda,\sigma}  , 
\eeq
with $\ol\Scf_{ij}$ representing the following combination:
\eq
\label\scfdef
\eqalign{
\ol\Scf_{ij} &\equiv \Re\mu_{ij} \; (\Ebar_x + \Bbar_y) + \Im\mu_{ij} \; (\Ebar_y -
\Bbar_x) \cr 
&= \Re\Bigl[ \mu_{ij} (\Ebar_x - i \Ebar_y) \Bigr] - \Im\Bigl[ \mu_{ij} 
(\Bbar_x - i \Bbar_y) \Bigr]. \cr
&= \Re\Bigl[ \mu_{ij} (\ol\Sce_x - i \ol\Sce_y) \Bigr] . \cr}
\eeq
Here $\Ebar_k$ and $\Bbar_l$ denote the (microscopic or ensemble) mean electric and
magnetic fields, averaged over the matter sector: 
\eq
\label\meanemfields
\Ebar_k = \Avgce{F_{0k}} \qquad \hbox{and} \qquad \Bbar_k = \hf \, \epsilon_{klm}
\Avgce{F_{lm}} ,
\eeq
and $\ol\Sce_k \equiv \Ebar_k + i \Bbar_k$. 

Similarly, the second-order contributions to the evolution equation become:
\eq
\label\mmvtwodiff
\eqalign{
(\Vbar_2)_{i\lambda;j\sigma} &= \delta_{\lambda\sigma} \int_{t'}^t d\tau \; 
\Bigl( \Avgce{\delta \Scf \delta \Scf} \Bigr)_{ij} \cr
\left( {\partial \rho^d_\ssf \over \partial t} \right)_{i\lambda;j\sigma} &= 
\hbox{sign}(\lambda) \, \hbox{sign}(\sigma) \, \int_{t'}^t d\tau \; \Bigl( 
\Avgce{\delta \Scf \rho_\ssf \delta \Scf} \Bigr)_{i\lambda;j\sigma} ,\cr}
\eeq
where $\delta\Scf_{ij} \equiv \Scf_{ij} - \ol\Scf_{ij}$. 

Clearly these expressions can be used to extend the analysis of fluctuations in
magnetic fields to systems for which the fluctuations are uncorrelated in momentum
space. Based on our experience with the MSW oscillations, for reasonably small
fluctuations we expect appreciable consequences dominantly in the presence of resonant
oscillations.

\section{Conclusions}

In this paper we have obtained the following results:

\topic{1}
We set up a general formalism for describing neutrino propagation
through fluctuating media. This formalism has the virtue that it is
derived from first principles, and so there are no hidden assumptions 
which limit its applications. As a result it can be applied to {\it 
any} source of fluctuations which may be of interest.

\topic{2} 
We have applied this formalism to many of the sources of fluctuations
for neutrinos propagating through the sun. When applied to microscopic,
thermal fluctuations we reproduce standard estimates, which give a
negligible impact on neutrino propagation. When applied to
larger-scale macroscopic density fluctuations the effects can be larger,
typically becoming important once $\epsilon^2 \ell \gsim 100$ m or so.
Here $\epsilon \sim \delta n/n$ is the relative amplitude of the 
density fluctuation, and $\ell$ is a measure of its correlation
length in the direction of neutrino motion. If this varies from
place to place, it is its value at the point where resonant oscillations
occur that is most important. 

\topic{3}
The neutrino evolution equation which we obtain --- eq.~\finalform\ --- agrees, when 
restricted to the domain of common validity, with those of previous workers 
\Equilib\Nusinbath\SiglRaffelt\Sawyer\noisy\noisytwo. 
We find the fluctuations found in ref.~\Sawyer\ to only significantly
decohere the neutrinos if their size is small compared to the size
of the detector. Since the strength of their influence on 
neutrino propagation is itself proportional to the fluctuation size,
such small fluctuations are likely to be negligible for solar neutrinos. 

\topic{4}
Two models of macroscopic solar fluctuations were developed. One
of these, having cells of constant, fluctuating density, 
reduces to ref.~\noisy, and is similar to ref.~\noisytwo, in the
limit of small cell size. The other is completely different, and
models the oscillatory density variations which occur in solar
acoustic $p$-waves. (A more realistic version of this model is currently
under study.) For these oscillatory waves, taking a period of 30 minutes, we
find appreciable neutrino effects for density fluctuations which
are at least a percent in size. We trace this comparatively small effect
to the relatively small wavelength at the neutrino resonance
point, which comes about because, for $p$ waves, the wavelength
decreases with depth due to the increase of the speed of sound.
We do not expect the same suppression to apply to $g$ waves.

\topic{5}
We integrate the resulting neutrino evolution equations for the
case of two neutrino flavours (with no CP violation) to obtain
a generalized Parke's formula --- eq.~\Parkeform\ --- for the electron-neutrino survival 
probability. This formula reproduces well our numerical integrations.
The decoherence due to the fluctuations
enters the neutrino evolution like a friction term, causing the
oscillations to damp. 

Two features emerge from the result. First, for small fluctuations 
deviations from the usual MSW survival probability first arise for
adiabatic resonant transitions, for which the MSW suppression 
deteriorates because the fluctuations partially ruin the resonance.
Because MSW oscillations use this suppression to agree with the 
solar neutrino data
by eliminating the ${}^7$Be line, measurements of this line in neutrino
detectors promises to shed light on the nature of density fluctuations
deep within the sun. 

Second, for small $E/\delta m^2$ and for sufficiently large fluctuations,
the survival probability falls to 0.5, independent of energy. This introduces
a new fluctuation-driven mechanism for solving the solar neutrino puzzle,
although its energy dependence is not favoured by current measurements.

\topic{6}
Finally, our formalism is used to derive the effective neutrino
hamiltonian which is relevant to magnetic moment couplings to
magnetic fields.  
\endtopic

\bigskip
\centerline{\bf Acknowledgements}
\bigskip

We would like to acknowledge Ira Rothstein for asking the question which
initiated this line of research, as well as fruitful discussions with 
Yulik Khriplovich, Mark Sutton, Bill Unruh and Nathan Weiss concerning coherence and
fluctuations within and outside of neutrino physics. Our research is financially
supported by NSERC of Canada and FCAR du Qu\'ebec.

\listrefs

\figurecaptions

\vfill\break
\bigskip\bigskip\bigskip\bigskip

\centerline{\epsfxsize=11.5cm\epsfbox[45 430 550 750]{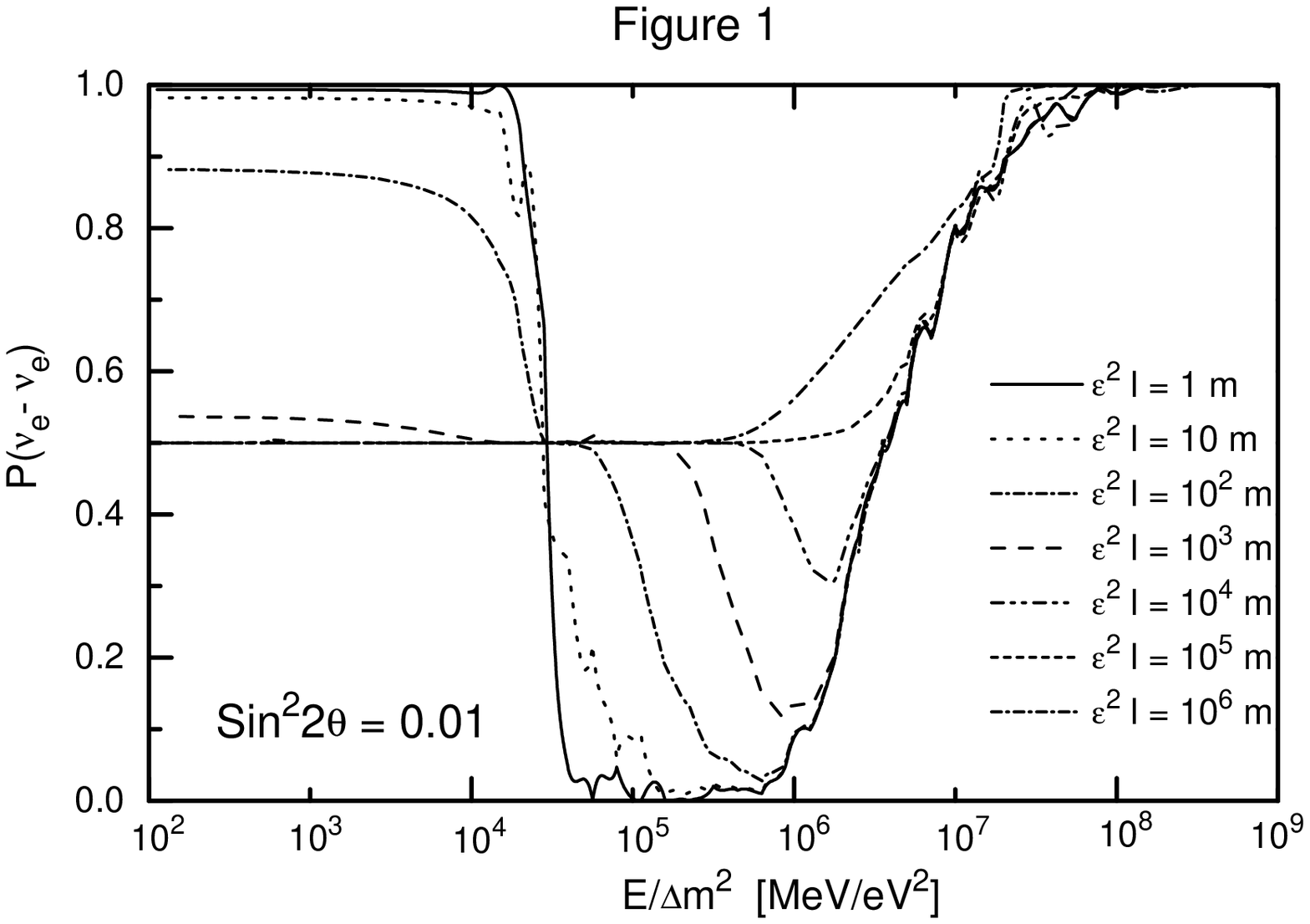}}
\bigskip\bigskip\bigskip\bigskip\bigskip
\centerline{\epsfxsize=11.5cm\epsfbox[45 430 550 750]{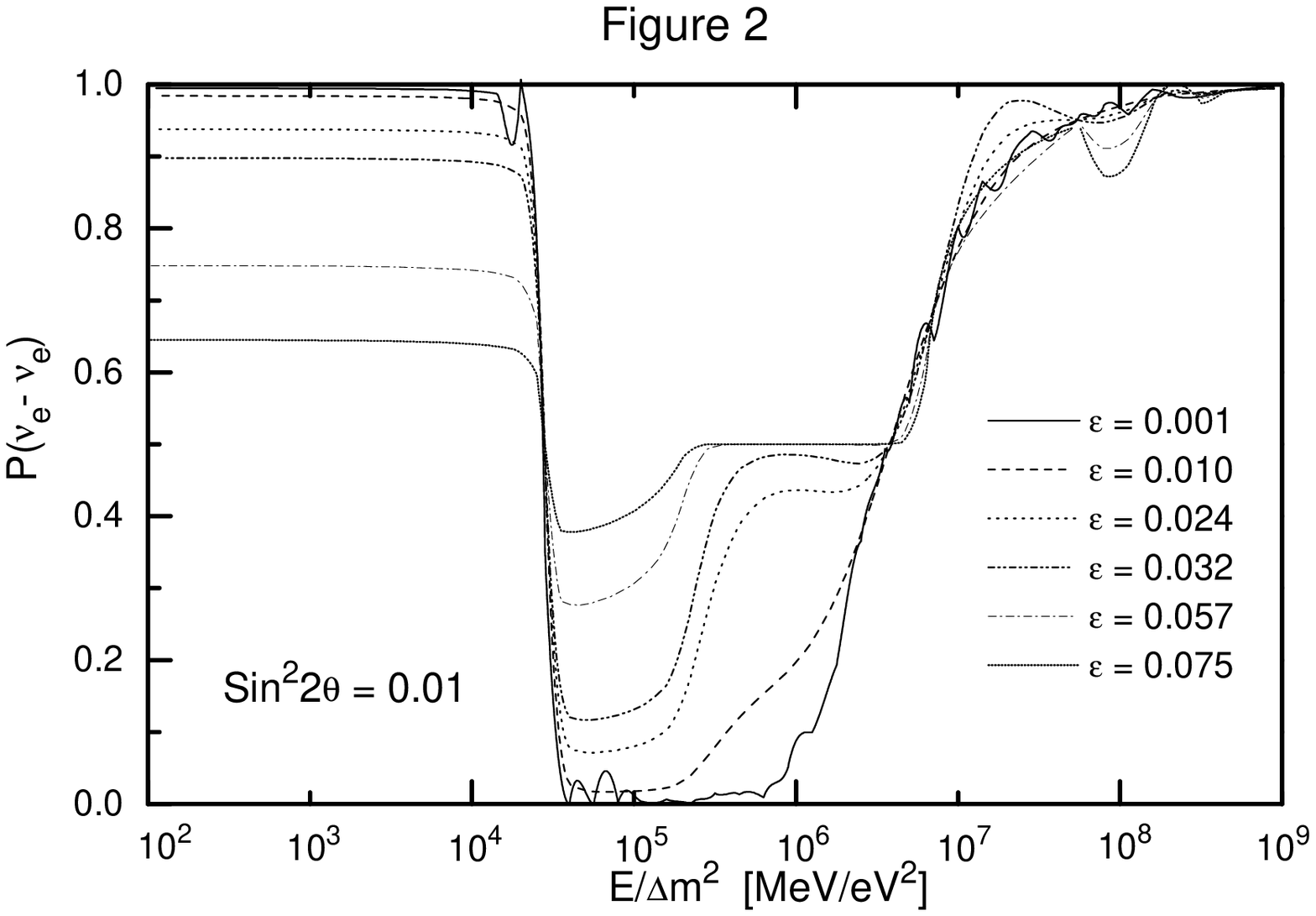}}

\bye